# Proof Analysis of A Foundational Classical Singlesuccedent Sequent Calculus

**Khashayar Irani**
**Birkbeck College, University of London**
**k.irani@mathematicallogic.com**
**July 2025**

## Abstract

In this paper we investigate the question: 'How can A Foundational Classical Singlesuccedent Sequent Calculus be formulated?'[1] The choice of this particular area of proof-theoretic study is based on a particular ground that is, to formulate a robust and foundational classical singlesuccedent sequent calculus that includes a number of novel rules with the ultimate aim of deriving the singlesuccedent sequent $\Gamma \Rightarrow C$.[2] To this end, we argue that among all standard sequent calculi (at least to the best of our knowledge) there is no classical singlesuccedent sequent calculus that can be considered the rightful successor to Gerhard Gentzen's (1935) original LK system. However, we also contend that while several classical singlesuccedent sequent calculi exist such as Sara Negri's and Jan von Plato's (2001 & 2011) G3ip+Gem-at and G0ip+Gem0-at calculi, none of these proof systems possess the classical proof-theoretic potential to meet the formal expectations of a dedicated classical proof theorist.[3] Conversely, we shall demonstrate that our forthcoming system, namely G-Calculus

---

[1] Initially, the full question addressed by this paper was: 'How can A Foundational Classical Singlesuccedent Sequent Calculus & its corresponding Classical Natural Deduction be formulated?' However, due to the considerable length of the paper and the potential issues this could cause for publication, we have divided the inquiry into two independent but related questions. The second part of this project, namely the segment concerning the corresponding classical natural deduction, shall follow after the publication of the present paper. Furthermore, we wish to emphasise that the aim of this paper (and its subsequent future segment) is modest. Our central objective in composing these manuscripts is to propose alternative classical proof systems, not only to enrich proof theory but also to support philosophical investigations into the foundations of logic.

[2] While the ultimate goal of this paper is to formulate a classical singlesuccedent sequent calculus capable of deriving the endsequent $\Gamma \Rightarrow C$ in all derivations, we shall also present at the end of this work the multisuccedent form of our proposed calculus whose end-sequent is $\Gamma \Rightarrow \Delta$ in order to demonstrate the power and flexibility of our classical scheme for future multisuccedent systems.

[3] In our view, a dedicated classical proof theorist expects a formal system to satisfy several rigorous standards, particularly when working within the confines of a singlesuccedent sequent calculus. As we shall see in section 3 of the paper, such a theorist seeks a system that not only upholds classical logical principles such as the law of the excluded middle and double negation elimination, but does so in a structurally harmonious and proof-theoretically robust manner. The system must allow for the derivation of sequents of the form $\Gamma \Rightarrow C$, while maintaining key metatheoretical properties such as the invertibility of logical rules and the admissibility of structural rules. Moreover, the presence of rules such as L¬ and R¬ is indispensable, as they facilitate symmetric reasoning and enable the application of reductio ad absurdum that is, a hallmark of classical logic. In this vein, atomic-level formulations (as seen in systems like G3ip+Gem-at) are regarded as insufficient, for they restrict classical reasoning to limited contexts. Instead, a fully expressive classical system must accommodate arbitrary

through its classical division i.e. Gc has been entirely designed to meet these expectations.[4] Prior to commencing our enquiry, a supplementary note must be made and that is in this work when discussing various sequent calculi, for proof-theoretic purposes, we are primarily concerned with their propositional components rather than their predicate divisions except in G-Calculus where we examine both aspects.

# 1 Proof Analysis of A Number of Eminent Classical Sequent Calculi

It is arguable that the most excellent, renowned and existing form of sequent calculus for classical logic is Gentzen's (1935) LK system.[5] In the literature of modern proof theory such as the works of Anne Sjerp Troelstra and Helmut Schwichtenberg (2000) or Negri and von Plato (2001) this style of sequent calculus is recognised as G1c for classical propositional and predicate calculus. The following proof system presents Gentzen's LK calculus including its axiom, logical and structural rules for propositional and predicate calculus with a number of minor notational modifications:[6]

**LK**

**Axiom**

---

formulae and preserve the syntactic and semantic integrity of classical derivations. Crucially, the system should ensure that logical operations can be applied uniformly through avoiding fragmentary treatment of complex formulae. As such, the proof theorist anticipates not merely a technically adequate system, but one that embodies the full proof-theoretical architecture of classical logic while being amenable to cut-elimination, modular proof construction, and goal-directed reasoning.

[4] Note that in this investigation the terms G-Calculus and Gc may occasionally be used interchangeably. This is because, the foundational classical singlesuccedent sequent calculus we aim to formulate is G-Calculus. However, we retain the label Gc to clearly distinguish the classical component of G-Calculus from its intuitionistic and minimal counterparts, namely the calculi Gi and Gm. Aside from drawing this distinction, G-Calculus and Gc refer to the same system.

[5] See Gentzen (1935) cited in M. E. Szabo (1969) Page: 83.

[6] There exist several notable distinctions between Gentzen's (1935) original LK system and the subsystem G1c each reflecting fundamental differences in their formulation and underlying structural principles. Firstly, in contrast to the operator ⊥, Gentzen employs his L¬ and R¬ rules as an alternative means of handling negation within the system. Secondly, while the LK system utilises sequences, G1c adopts multisets, a difference that necessitates the introduction of the structural rule of interchange (Inc) also referred to as permutation in Gentzen's framework to allow for the rearrangement of formulae within derivations. Thirdly, unlike G1c, Gentzen's formulation of the L→ rule does not include a sharing context thus enforcing a stricter structural constraint on the application of implications within proofs. Fourthly, in his presentation of the LK system Gentzen includes the cut rule as an integral part of the system, whereas in G1c the system is formulated without the explicit addition of the cut rule demonstrating a different approach to proof structuring and the elimination of intermediary steps. For a more detailed discussion of these differences see Gentzen (1935) cited in Szabo (1969) Page: 83.

**A ⇒A**

**Logical Rules**

| A,Γ ⇒ Δ | B,Γ ⇒ Δ | Γ ⇒ Δ,A    Γ ⇒ Δ,B |
|---|---|---|
| ——————————L∧ | ——————————L∧ | ————————————————R∧ |
| A ∧ B,Γ ⇒ Δ | A ∧ B,Γ ⇒ Δ | Γ ⇒ Δ,A ∧ B |

| A,Γ ⇒ Δ    B,Γ ⇒ Δ | Γ ⇒ Δ,A | Γ ⇒ Δ,B |
|---|---|---|
| ——————————————L∨ | ——————————R∨ | ——————————R∨ |
| A ∨ B,Γ ⇒ Δ | Γ ⇒ Δ,A ∨ B | Γ ⇒ Δ,A ∨ B |

| Γ ⇒ Δ,A    B.Θ ⇒ Λ | A.Γ ⇒ Δ,B |
|---|---|
| ——————————————L→ | ——————————————R→ |
| A → B.Γ,Θ ⇒ Δ,Λ | Γ ⇒ Δ,A → B |

| Γ ⇒ Δ,A | A,Γ ⇒ Δ |
|---|---|
| ——————————L¬ | ——————————R¬ |
| ¬A,Γ ⇒ Δ | Γ ⇒ Δ,¬A |

| A(x/t),Γ ⇒ Δ | Γ ⇒ Δ,A(x/y) |
|---|---|
| ——————————L∀ | ——————————R∀ |
| ∀xAx,Γ ⇒ Δ | Γ ⇒ Δ,∀xAx |

| A(x/y),Γ ⇒ Δ | Γ ⇒ Δ,A(x/t) |
|---|---|
| ——————————L∃ | ——————————R∃ |
| ∃xAx,Γ ⇒ Δ | Γ ⇒ Δ,∃xAx |

**Structural Rules**

| Γ ⇒ Δ | Γ ⇒ Δ |
|---|---|
| ————————LW | ————————RW |
| A,Γ ⇒ Δ | Γ ⇒ Δ,A |

| A,A,Γ ⇒ Δ | Γ ⇒ Δ,A,A |
|---|---|
| ——————————LC | ——————————RC |
| A,Γ ⇒ Δ | Γ ⇒ Δ,A |

| Θ,B,C,Γ ⇒ Δ | Γ ⇒ Δ,C,B,Λ |
|---|---|
| ——————————————LInc | ——————————————RInc |
| Θ,C,B,Γ ⇒ Δ | Γ ⇒ Δ,B,C,Λ |

**Γ ⇒ Δ,A    A,Θ ⇒ Λ**

_____________________**Cut**
$\Gamma,\Theta \Rightarrow \Delta,\Lambda$

Prior to continuing with our proof analysis, we should emphasise a significant matter regarding the presentation of predicate calculus' formalism, namely the distinction between bound or proper and free variables is crucial in this calculus. This is because, the formulae presented demonstrate how the variable y is bound or forms a proper formula, whereas t remains free. In the rule $\forall$I, a variable must be arbitrarily chosen and not free in any undischarged assumption. The formula A(x/y) represents a generalisation where y is a bound variable because its instantiation is arbitrary and does not depend on any particular term. When we conclude $\forall$xAx the variable x is now universally quantified i.e. it is formally bound within the scope of the quantifier. Conversely, in the rule $\forall$E, the instantiation A(x/t) involves substituting the quantified variable x with a specific term t. However, t is not quantified within the proof making it a free term rather than a bound variable. Equally, in the rule $\exists$I, replacing x with a particular term t allows us to assert that there exists an x for which Ax holds making t a free instance of the predicate. But in the rule $\exists$E, a variable y is introduced as a bound variable to generalise the assumption A(x/y) ensuring that no dependency on any specific term t remains. Thus, y is bound when introduced arbitrarily in $\forall$I and $\exists$E, whereas t remains free when instantiated in $\forall$E and $\exists$I. Nonetheless, leaving these predicate calculi matters aside, we argue while Gentzen's LK is a forceful and classical formal proof system, but since our primary objective is to formulate a classical singlesuccedent sequent calculus system capable of deriving the sequent $\Gamma \Rightarrow C$, therefore LK does not serve our purpose. A closer examination of contemporary works in proof analysis may provide a solution to this problem. In this regard, Negri's and von Plato's (2001) G3ip+Gem-at calculus appears to be an excellent choice.[7] This classical singlesuccedent sequent calculus system is moderately a meticulous proof system for both classical propositional and predicate calculus.[8] The system G3ip+Gem-at is derived by adding the sequent form of the rule of excluded middle (LEM):

$P.\Gamma \Rightarrow C$ $\quad\quad\quad\quad$ $\neg P.\Gamma \Rightarrow C$
___________________________**Gem-at**
$\Gamma \Rightarrow C$

for atomic formulae to the intuitionistic system G3ip.[9] The rule Gem-at permits for any atomic formula P, a derivation of $\Gamma \Rightarrow C$ from derivations of both P,$\Gamma \Rightarrow C$ and $\neg$P.$\Gamma \Rightarrow C$.[10]

_______________________________

[7] Negri & von Plato (2001) Page: 114

[8] Negri & von Plato (2011) Pages: 92 & 222-224 argue that, from a proof-theoretic point of view, the principal rationale for introducing a singlesuccedent calculus such as G3ip+Gem-at is to extend the operational interpretation of sequents to include classical propositional calculus, while simultaneously guaranteeing the preservation of decidability within intuitionistic predicate proof theory. In classical predicate calculus, the decidability of a quantified formula such as ∀xAx or ∃xAx cannot, in general, be inferred from the decidability of its individual instances A(x/t) for the arbitrary term t. Nevertheless, by including suitably formulated quantifier rules e.g. modelled after the treatment of quantifiers in systems such as Heyting Arithmetic into the structure of G3ip+Gem-at, it becomes possible to establish a decidable system of classical predicate logic.

[9] In the calculus G3ip+Gem-at the rule Gem-at for atoms implies a general form of reductio ad absurdum for atoms in natural deduction.

A fundamental characteristic of this system is that it is able to satisfy the admissibility of the structural rules of weakening, contraction and cut.[11] It is important to note that in G3ip+Gem-at, LEM is initially introduced for atomic formulae, not for arbitrary formulae:

$$\frac{A.\Gamma \Rightarrow C \qquad \neg A.\Gamma \Rightarrow C}{\Gamma \Rightarrow C} \text{Gem}$$

Accordingly, G3ip+Gem-at cannot immediately become a fully classical system for propositional or predicate calculus.[12] However, while G3ip+Gem-at at the propositional level initially ensures LEM only for atomic formulae, the system is designed such that, upon the addition of suitably formulated quantifier rules that is, modelled on the treatment of quantifiers in intuitionistic systems like Heyting Arithmetic, it becomes possible to establish a decidable classical predicate calculus where LEM holds for arbitrary formulae.[13] Thus, although the base system G3ip+Gem-at enforces Gem at the atomic level, its extension by intuitionistic-style quantifiers achieves LEM for all formulae, thus ensuring a full classical logical framework. A further characteristic of the calculus G3ip+Gem-at is that it constitutes a well-behaved proof system as all intuitionistic constants are present within it.[14] Although LEM is employed for atoms, derivations nevertheless remain cut-free. As a result, the subformula property is maintained for cut-free derivations. This ensures that proofs remain syntactically disciplined, as no extraneous formulae beyond subformulae of the endsequent are introduced during derivations without cuts. Conversely, since the calculus G3ip+Gem-at is based on G3ip, namely a form of intuitionistic sequent calculus, consequently it lacks fundamental classical rules such as L¬ and R¬. Hence, it follows that to fulfil our classical proof-theoretic obligations, we require a form of classical singlesuccedent sequent calculus that both retains the advantageous properties of G3ip+Gem-at and includes classical principles such as L¬, R¬ and possibly Gem-at or Gem.

---

[10] See Negri & von Plato (2011) Page: 93. Furthermore, as Negri & von Plato (2011) Page: 94 argue, in terms of decidability, while G3ip+Gem-at introduces a restricted classical element through Gem-at, it does not achieve decidability for predicate logic. Similar to first order logic, intuitionistic predicate classical logic remains undecidable even with the addition of the rule Gem-at. Therefore, G3ip+Gem-at does not present a decidable first order predicate logic system.

[11] Negri & von Plato (2011) Page: 92

[12] Negri & von Plato (2011) Page: 93

[13] See Negri & von Plato (2001) Page: 160 and Negri & von Plato ( 2011) Pages: 92 & 222–224.

[14] Negri & von Plato (2001) Page: 160 and Negri & von Plato (2011) Pages: 92-93 suggest that the singlesuccedent structure of the system G3ip+Gem-at is designed to support a well-behaved operational reading of sequents, preserving a close correspondence with natural deduction systems. However, the addition of rules such as Gem-at and Gem which introduce classical reasoning through LEM complicates the direct translation into a singleconclusion natural deduction system. While G3ip on its own lends itself naturally to such a translation, the structural consequences of Gem-at and Gem require their removal if a strictly intuitionistic, singleconclusion natural deduction form is to be achieved. This interpretation is grounded in their discussion of the structural impact of classical principles on proof systems and the operational motivations for preferring singlesuccedent frameworks.

It is important to note that within the family of sequent calculi there exists a fundamental yet powerful form of singlesuccedent sequent calculus known as G0.[15] The number "0" refers to the fact that the G0 calculus is the preliminary sequent calculus and all other sequent calculi are built on the G0. The calculus G0 in its basic form i.e. in minimal logic for the propositional calculus can be demonstrated as:[16]

**G0m**

**Axiom**

A ⇒ A

**Logical Rules**

$$\frac{A,B,\Gamma \Rightarrow C}{A \wedge B,\Gamma \Rightarrow C}L\wedge \qquad \frac{\Gamma \Rightarrow A \quad \Gamma \Rightarrow B}{\Gamma \Rightarrow A \wedge B}R\wedge$$

$$\frac{A,\Gamma \Rightarrow C \quad B,\Gamma \Rightarrow C}{A \vee B,\Gamma \Rightarrow C}L\vee \qquad \frac{\Gamma \Rightarrow A}{\Gamma \Rightarrow A \vee B}R\vee \qquad \frac{\Gamma \Rightarrow B}{\Gamma \Rightarrow A \vee B}R\vee$$

$$\frac{\Gamma \Rightarrow A \quad B.\Gamma \Rightarrow C}{A \rightarrow B.\Gamma \Rightarrow C}L\rightarrow \qquad \frac{A.\Gamma \Rightarrow B}{\Gamma \Rightarrow A \rightarrow B}R\rightarrow$$

**Structural Rules**

$$\frac{\Gamma \Rightarrow C}{A,\Gamma \Rightarrow C}LW \qquad \frac{A.A,\Gamma \Rightarrow C}{A,\Gamma \Rightarrow C}LC$$

On the other hand, if we add the principle L⊥:

$$\frac{}{\perp \Rightarrow}L\perp$$

to G0mp, we will obtain G0ip which is an intuitionistic singlesuccedent sequent calculus. The calculus G0ip improves upon G3ip by eliminating redundancies in the formulation of rules, maintaining invertibility, as well as subformula property. A central point about the calculi

---

G0mp and G0ip should be made and that is, the cut rule is not present in these systems because it is eliminable.[17] However, if we go further and add the rule Gem0-at:

$$\frac{P.\Gamma \Rightarrow C \qquad\qquad \neg P.\Gamma \Rightarrow C}{\Gamma \Rightarrow C}\text{Gem0-at}$$

to G0ip, then we will acquire G0ip+Gem0-at or in other words G0cp which is a classical singlesuccedent sequent calculus. The addition of the rule Gem0-at, like Gem-at, grants atomic excluded middle to G0ip by permitting a derivation of $\Gamma \Rightarrow C$ from both $P.\Gamma \Rightarrow C$ and $\neg P.\Gamma \Rightarrow C$, where P is atomic. This inclusion of atomic excluded middle again restricts classical reasoning to the atomic level through preserving the underlying intuitionistic framework for complex formulae. As with G3ip+Gem-at, the structural rules of weakening, contraction, and cut remain height-preservingly admissible in G0ip+Gem0-at.[18] However, as in G3ip+Gem-at, cut elimination in the strictest sense is only secured for cuts on atomic formulae. Cut elimination for arbitrary complex formulae cannot be guaranteed due to the integration of atomic excluded middle.[19] Nonetheless, the system maintains the subformula property for derivations without cuts, an important syntactic discipline guaranteeing that proofs do not expand beyond the logical content of the original endsequent. The important observation is that G0ip+Gem0-at retains the intuitionistic discipline at the level of compound formulae, even as it admits classical reasoning for atomic propositions. Thus, G0ip+Gem0-at provides a structure similar to G3ip+Gem-at but offers a more streamlined approach to derivation construction due to the syntactic simplifications inherent in G0ip. Additionally, proof search within G0ip+Gem0-at remains systematic and controlled by benefiting from the invertibility of logical rules and the admissibility of structural transformations.

Furthermore, it is very important to correct any mischaracterisation suggesting that G3ip+Gem-at or G0ip+Gem0-at are complete systems for classical predicate calculus.[20] The extensions Gem-at and Gem0-at introduce classical reasoning only at the atomic level. They do not allow derivation of LEM for arbitrary complex formulae. Thus, G3ip+Gem-at and G0ip+Gem0-at are not complete for classical predicate logic and must be viewed as primarily intuitionistic systems with localised classical augmentation at the atomic level. This hybrid character situates both systems uniquely within the landscape of sequent calculi. They permit limited classical reasoning without compromising the core structure and advantages of intuitionistic proof theory. Their primary proof-theoretic virtues  e.g. invertibility of rules, subformula property and admissibility of structural rules  remain intact despite the introduction of atomic excluded middle. Thus, they exemplify how classical elements can be carefully integrated into an intuitionistic framework without sacrificing proof-theoretic discipline.[21] Moreover, when considering their applications, G3ip+Gem-at and G0ip+Gem0-at are particularly suited for computational interpretations of logic. In areas such as automated proof search, logic programming, and constructive type theory, it is often advantageous to reason classically about atomic facts while retaining constructivist discipline for more complex constructions. The systems' proof-theoretic features  such as controlled use of classicality, preservation of subformula property and invertibility of logical rules  make them highly compatible with such computational frameworks.[22] Finally, while the two systems G3ip+Gem-at and G0ip+Gem0-at are proof-theoretically equivalent in their logical strength and behaviour, G0ip+Gem0-at is syntactically more efficient. Its rules are formulated to eliminate redundancy and minimise unnecessary complexity in derivations.[23] Thus, while G3ip+Gem-at may offer a more direct extension of traditional intuitionistic sequent calculi, G0ip+Gem0-at represents an evolution towards greater syntactic economy by making it particularly attractive for foundational studies and applications where proof efficiency is paramount.

---

[17] Negri & von Plato (2011) Page: 93
[18] Negri & von Plato (2011) Page: 93
[19] Negri & von Plato (2011) Page: 93
[20] Negri & von Plato (2011) Pages: 93-94

In summary, the systems G3ip+Gem-at and G0ip+Gem0-at must be understood as structurally sound intuitionistic sequent calculi extended with a carefully restricted form of classical reasoning at the atomic level. They maintain the essential proof-theoretic properties of height-preserving admissibility of structural rules, invertibility of logical rules, and the subformula property for cut-free derivations. However, they do not achieve completeness for classical predicate logic and should not be mistaken for fully classical systems. Instead, they offer a hybrid proof-theoretic architecture that strategically integrates classical reasoning while preserving the constructive rigour and syntactic discipline characteristic of intuitionistic logic. Nonetheless, if we set aside the proof-theoretic elegance of G3ip+Gem-at and G0ip+Gem0-at, since these two calculi lack the two classical rules of L¬ and R¬, they cannot satisfy our classical expectations, namely the formulation of a foundational singlesuccedent LK-style sequent calculus.

## 2 The Preamble of Three Nonstandard Classical Sequent Calculi

In this section, we briefly refer to Andrzej Indrzejczak's (2014) paper *A Survey of Nonstandard Sequent Calculi* in which he provides a comprehensive review of various sequent calculi that deviate from the standard framework introduced by Gentzen.[24] However,

---

[21] In addition to these technical attributes, the philosophical implications of G3ip+Gem-at and G0ip+Gem0-at are noteworthy. By restricting classical reasoning to the atomic level, these systems embody a cautious and selective approach to classicality, one that preserves the constructive content of intuitionistic logic for compound propositions while allowing atomic cases to behave classically. This reflects a broader trend in proof theory towards exploring intermediate systems that balance the demands of constructivism and classical logic.

[22] Negri & von Plato (2001) Page: 157

[23] Negri & von Plato (2011) Page: 93

[24] In this survey Indrzejczak categorises these calculi based on their structural and rule-based deviations from the classical form, aiming to highlight alternative proof systems that have been developed for theoretical and practical advancements in logic. The paper begins by contextualising the notion of sequent calculus (SC) traditionally associated with Gentzen's formalism in which structural and logical rules dictate how logical constants are introduced or manipulated in a proof. However, modern research has expanded this framework to

prior to present Indrzejczak's proof-theoretic study we should state that, although in demonstrating his sequent calculus formulations we have made slight notational alterations e.g. replacing Greek letters with English ones, the formal structures of the systems in question have been fully preserved. One of the classical systems Indrzejczak discusses is Gentzen's NK in sequent calculus style (SND).[25] The following formalism is Indrzejczak's formulation of Gentzen's NK for propositional and predicate calculus in sequent-style:[26]

**Axiom**

$A \Rightarrow A$

**Logical Rules**

$$\frac{\Gamma \Rightarrow A \quad \Delta \Rightarrow B}{\Gamma,\Delta \Rightarrow A \wedge B} \Rightarrow \wedge I \qquad \frac{\Gamma \Rightarrow A \wedge B}{\Gamma \Rightarrow A} \Rightarrow \wedge E \qquad \frac{\Gamma \Rightarrow A \wedge B}{\Gamma \Rightarrow B} \Rightarrow \wedge E$$

$$\frac{\Gamma \Rightarrow A}{\Gamma \Rightarrow A \vee B} \Rightarrow \vee I \qquad \frac{\Gamma \Rightarrow B}{\Gamma \Rightarrow A \vee B} \Rightarrow \vee I \qquad \frac{\Gamma,A \Rightarrow C \quad \Delta,B \Rightarrow C \quad \Pi \Rightarrow A \vee B}{\Gamma,\Delta,\Pi \Rightarrow C} \Rightarrow \vee E$$

$$\frac{\Gamma,A \Rightarrow B}{\Gamma \Rightarrow A \to B} \Rightarrow \to I \qquad \frac{\Gamma \Rightarrow A \quad \Delta \Rightarrow A \to B}{\Gamma,\Delta \Rightarrow B} \Rightarrow \to E$$

$$\frac{\Gamma,A \Rightarrow B \quad \Delta,A \Rightarrow \neg B}{\Gamma,\Delta \Rightarrow \neg A} \Rightarrow \neg I \qquad \frac{\Gamma \Rightarrow \neg\neg A}{\Gamma \Rightarrow A} \Rightarrow \neg\neg E$$

$$\frac{\Gamma \Rightarrow A(x/y)}{\Gamma \Rightarrow \forall x Ax} \Rightarrow \forall I \qquad \frac{\Gamma \Rightarrow \forall x Ax}{\Gamma \Rightarrow A(x/t)} \Rightarrow \forall E$$

| | | | |
|---|---|---|---|
| $\Gamma \Rightarrow A(x/t)$ | | $\Gamma \Rightarrow \exists xAx \quad A(x/y),\Delta \Rightarrow B$ | |
| ___________ $\Rightarrow\exists I$ | | _______________________ $\Rightarrow\exists E$ | |
| $\Gamma \Rightarrow \exists xAx$ | | $\Gamma,\Delta \Rightarrow B$ | |

The above formalism has been modified by restricting the introduction of logical constants to the succedent. This suggests instead of using rules to introduce assumptions into the antecedent, it employs elimination rules in the succedent. This results in a system that combines elements of natural deduction and sequent calculus leading to proof structures that closely resemble those found in natural deduction but with explicit sequencing of premises. This system was initially formulated to make possible the proof of consistency in Peano Arithmetic and it is distinct from Gentzen's NK which operates directly on formulae rather than sequents. In natural deduction in sequent-style, all inference rules apply to sequents rather than formulae making it an intermediary between standard sequent calculus e.g. LK and natural deduction. One significant feature of such a method is its elimination of explicit assumption discharge mechanisms which are prevalent in traditional natural deduction. Instead, it maintains a sequential structure where all assumptions are carried forward through the derivation. This approach simplifies bookkeeping in proofs by avoiding nested subproofs a feature commonly associated with Jaśkowski's style of natural deduction. Furthermore, the natural deduction rules in sequent calculus style operate exclusively on succedents which means that introduction rules for logical constants follow a progressive structure. Logical constants are introduced only in the conclusion while their elimination rules extract inferential consequences from established sequents. This design maintains a close connection to the subformula property ensuring that derived sequents remain closely tied to their premises. Moreover, this process also allows for a more linear representation of proofs compared to standard tree-like structures. This linearisation facilitates proof search and mechanisation as each inference step explicitly preserves all active assumptions without requiring nested derivations. However, this benefit comes at the cost of redundancy as all assumptions must be restated at every inference step. This results in longer proofs compared to conventional sequent calculi where assumptions can be discharged more flexibly. A crucial derivable rule in this system is the cut rule which follows naturally from the way implications and negations are handled. Unlike standard LK where the cut rule is a primitive structural rule, natural deduction in sequent-style derives cut through its systematic handling of assumptions and logical rules. This derivability emphasises the system's foundational lucidity as it can simulate key inferential mechanisms found in both natural deduction and sequent calculus.[27]

Nevertheless, despite some of its proof-theoretic advantages e.g. its classicality and metamathematical uses, such a formal structure cannot fulfil our objective of deriving the singlesuccedent sequent $\Gamma \Rightarrow C$. This is because, in order to convert natural deduction rules into sequent-style principles, all rules governing logical constants have been exclusively introduced in the succedent. As a result, rather than employing rules to introduce assumptions into the antecedent, it relies on elimination rules in the succedent. The central problem with this approach is that it entirely disrupts the standard symmetry that exists between right and left rules in sequent calculus. As we shall see in the next section where we put forward our proposed sequent calculus apparatus, symmetry establishes an isomorphic relationship between right and left rules enabling us to derive the endsequent $\Gamma \Rightarrow C$ by applying a cut rule to the relevant right and left rules. For example, such an isomorphic relationship between the rules of sequent calculus can be observed in the rules governing $\wedge$ and $\vee$:

$$\frac{\Gamma \Rightarrow A \quad \Gamma \Rightarrow B}{\Gamma \Rightarrow A \wedge B}R\wedge \qquad \frac{A,B,\Gamma \Rightarrow C}{A \wedge B,\Gamma \Rightarrow C}L\wedge$$

$$\frac{}{\Gamma \Rightarrow C}Cut$$

$$\frac{\Gamma \Rightarrow A}{\Gamma \Rightarrow A \vee B}R\vee \qquad \frac{A,\Gamma \Rightarrow C \quad B,\Gamma \Rightarrow C}{A \vee B,\Gamma \Rightarrow C}L\vee$$

$$\frac{}{\Gamma \Rightarrow C}Cut$$

---

[27] Despite its theoretical appeal, Gentzen's natural deduction system in sequent calculus style is rarely used in practical proof search due to its verbosity. Each inference step necessitates the explicit rewriting of all assumptions making the process cumbersome for large-scale derivations. Nevertheless, its structure has influenced various modifications and refinements particularly in systems that aim to bridge natural deduction and sequent calculus while maintaining a linear proof format.

Now consider the situation in which we wished to derive the endsequent $\Gamma,\Delta,\Pi \Rightarrow C$ from the isomorphic relation that exists between the rules $\Rightarrow\vee I$ and $\Rightarrow\vee E$:

$$\frac{\Gamma \Rightarrow A}{\Gamma \Rightarrow A \vee B}\Rightarrow\vee I \qquad \frac{\Gamma \Rightarrow B}{\Gamma \Rightarrow A \vee B}\Rightarrow\vee I \qquad \frac{\Gamma,A \Rightarrow C \quad \Delta,B \Rightarrow C \quad \Pi \Rightarrow A \vee B}{\Gamma,\Delta,\Pi \Rightarrow C}\Rightarrow\vee E$$

However, this derivation cannot be achieved because these two rules are not symmetrically formulated and therefore the cut rule cannot be applied to them. Next, another motivating classical natural deduction in sequent-style system that Indrzejczak talks about in his comprehensive survey belongs to Hans Hermes (1963) in which he extends this framework by permitting logical operations in the antecedent. This relaxation enables a more flexible proof search process. Hermes' system can be presented for propositional and predicate calculus in the following manner:[28]

**Axioms**

$A \Rightarrow A \qquad \Rightarrow t = t$

**Logical Rules**

[28] Indrzejczak (2014) Page: 1301

$$\frac{\Gamma \Rightarrow A \quad \Delta \Rightarrow B}{\Gamma,\Delta \Rightarrow A \wedge B} \Rightarrow \wedge I \qquad \frac{\Gamma \Rightarrow A \wedge B}{\Gamma \Rightarrow A} \Rightarrow \wedge E \qquad \frac{\Gamma \Rightarrow A \wedge B}{\Gamma \Rightarrow B} \Rightarrow \wedge E$$

$$\frac{\Gamma \Rightarrow A}{\Gamma \Rightarrow A \vee B} \Rightarrow \vee I \qquad \frac{\Gamma \Rightarrow B}{\Gamma \Rightarrow A \vee B} \Rightarrow \vee I \qquad \frac{\Gamma,A \Rightarrow C \quad \Delta,B \Rightarrow C}{\Gamma,\Delta,A \vee B \Rightarrow C} \Rightarrow \vee E$$

$$\frac{\Gamma,A \Rightarrow B}{\Gamma \Rightarrow A \rightarrow B} \Rightarrow \rightarrow I \qquad \frac{\Gamma,\neg A \Rightarrow C \quad \Delta,B \Rightarrow C}{\Gamma,\Delta,A \rightarrow B \Rightarrow C} \Rightarrow \rightarrow E$$

$$\frac{\Gamma,A \Rightarrow B \quad \Delta,\neg A \Rightarrow B}{\Gamma,\Delta \Rightarrow B} \Rightarrow \neg E \qquad \frac{\Gamma \Rightarrow A \quad \Delta \Rightarrow \neg A}{\Gamma,\Delta \Rightarrow B} \Rightarrow \neg E$$

$$\frac{\Gamma \Rightarrow A(x/y)}{\Gamma \Rightarrow \forall x Ax} \Rightarrow \forall I \qquad \frac{\Gamma \Rightarrow \forall x Ax}{\Gamma \Rightarrow A(x/t)} \Rightarrow \forall E$$

$$\frac{\Gamma \Rightarrow A(x/t)}{\Gamma \Rightarrow \exists x Ax} \Rightarrow \exists I \qquad \frac{\Gamma,A(x/y) \Rightarrow B}{\Gamma,\exists x Ax \Rightarrow B} \Rightarrow \exists E$$

$$\frac{\Gamma \Rightarrow A}{\Gamma,x = t \Rightarrow A(x/t)} \Rightarrow = \qquad \frac{\Gamma \Rightarrow A}{\Gamma' \Rightarrow A'} \text{Subst}$$

Hermes' system includes structural rules such as weakening, contraction and permutation but also introduces direct elimination rules for negation and quantification. This approach makes the proof system more adaptable while preserving core logical properties such as cut elimination. Nonetheless, although Hermes' system is similar to Gentzen's formalism yet it fails to meet our proof-theoretic expectations. This is because, aside from lacking a proper $\Rightarrow\neg I$ rule, it is also unable to derive our desired single-succedent sequent $\Gamma \Rightarrow C$ due to the same issues encountered in Gentzen's system. Therefore, we should set aside Hermes' construction and seek a more suitable sequent-based system. Eventually, an interesting and powerful classical sequent calculus system included in Indrzejczak's review deserving our attention is Raymond Smullyan's (1968) multisuccedent sequent calculus which includes various left and right negation rules. The following formal scheme is Smullyan's system for propositional and predicate calculus:[29]

---

[29] Indrzejczak (2014) Page: 1306

**Logical Rules**

$$\frac{\Gamma,\neg A \Rightarrow \Delta \quad \Gamma,\neg B \Rightarrow \Delta}{\Gamma,\neg(A \wedge B) \Rightarrow \Delta} L\neg\wedge \qquad \frac{\Gamma \Rightarrow \Delta,\neg A,\neg B}{\Gamma \Rightarrow \Delta,\neg(A \wedge B)} R\neg\wedge$$

$$\frac{\Gamma,\neg A,\neg B \Rightarrow \Delta}{\Gamma,\neg(A \vee B) \Rightarrow \Delta} L\neg\vee \qquad \frac{\Gamma \Rightarrow \Delta,\neg A \quad \Gamma \Rightarrow \Delta,\neg B}{\Gamma \Rightarrow \Delta,\neg(A \vee B)} R\neg\vee$$

$$\frac{\Gamma,A,\neg B \Rightarrow \Delta}{\Gamma,\neg(A \rightarrow B) \Rightarrow \Delta} L\neg\rightarrow \qquad \frac{\Gamma \Rightarrow \Delta,A \quad \Gamma \Rightarrow \Delta,\neg B}{\Gamma \Rightarrow \Delta,\neg(A \rightarrow B)} R\neg\rightarrow$$

$$\frac{\Gamma,\neg A \Rightarrow \Delta \quad \Gamma,B \Rightarrow \Delta}{\Gamma,A \rightarrow B \Rightarrow \Delta} L\rightarrow \qquad \frac{\Gamma \Rightarrow \Delta,\neg A,B}{\Gamma \Rightarrow \Delta,A \rightarrow B} R\rightarrow$$

$$\frac{\Gamma,A \Rightarrow \Delta}{\Gamma,\neg\neg A \Rightarrow \Delta} L\neg\neg \qquad \frac{\Gamma \Rightarrow \Delta,A}{\Gamma \Rightarrow \Delta,\neg\neg A} R\neg\neg$$

$$\frac{\Gamma,\forall xAx,A(x/t) \Rightarrow \Delta}{\Gamma,\forall xAx \Rightarrow \Delta} L\forall \qquad \frac{\Gamma \Rightarrow \Delta,\exists xAx,A(x/t)}{\Gamma \Rightarrow \Delta,\exists xAx} R\exists$$

$$\frac{\Gamma,\exists x\neg Ax \Rightarrow \Delta}{\Gamma,\neg\forall xAx \Rightarrow \Delta} L\neg\forall \qquad \frac{\Gamma \Rightarrow \Delta,\exists x\neg Ax}{\Gamma \Rightarrow \Delta,\neg\forall xAx} R\neg\forall$$

$$\frac{\Gamma,\forall x\neg Ax \Rightarrow \Delta}{\Gamma,\neg\exists xAx \Rightarrow \Delta} L\neg\exists \qquad \frac{\Gamma \Rightarrow \Delta,\forall x\neg Ax}{\Gamma \Rightarrow \Delta,\neg\exists xAx} R\neg\exists$$

Smullyan's classical multisuccedent sequent calculus is particularly designed for proving interpolation theorems. This system modifies the standard Gentzen-style sequent calculus by enforcing a strict separation of introduction rules into two distinct categories, namely those that operate in the antecedent and those that operate in the succedent. Crucially, the Smullyan system eliminates rules that permit the transfer of formulae from one side of a sequent to the other which significantly impacts the treatment of negation and implication. In contrast to Gentzen's sequent calculus, Smullyan's system lacks standard introduction rules for negation. Instead, it introduces negated complex formulae via a dedicated pair of rules which also allows for the introduction of double negation. This is a fundamental departure from traditional sequent calculus where negation is typically introduced and manipulated through structural transformations. The absence of formula transfer between antecedents and succedents necessitates an alternative approach to handling negation. Furthermore, Smullyan's calculus modifies Gentzen's treatment of implication. The standard introduction and elimination rules for implication are replaced by rules tailored to accommodate the system's constraints on formula transfer. These modifications ensure that negation and implication operate within the strictly defined structure of antecedent and succedent specific rules. Moreover, the formalisation of predicate calculus within the Smullyan's system is achieved through the inclusion of eight additional rules. Among these, the rules for universal quantification and existential quantification follow standard Gentzen-style formulations. However, the remaining rules reflect the unique properties of the system ensuring that all logical constants are treated in a way that preserves the strict delineation between antecedent and succedent operations. On the other hand, while Smullyan's system is a well-structured classical sequent calculus that fully includes the fundamental aspects of classical logic such as symmetry and negation rules and is arguably superior to any other classical system discussed thus far (with the exception of Gentzen's LK), it remains a multisuccedent system. Consequently, it does not satisfy our initial proof-theoretic expectation, namely the ability to derive the single-succedent sequent $\Gamma \Rightarrow C$. Therefore, we are compelled to exclude this formalism as we have with the previous ones and proceed to the next section where we present our proposed solution.[30]

# 3 G-Calculus: A Foundational Classical Singlesuccedent Sequent Calculus

Arguably, the proof system G-Calculus should not be assigned a number since historically speaking, although Gentzen never explicitly discussed this system, if he had it would have preceded the current G0 systems of the previous section and would in some sense constitute the singlesuccedent form of his LK system. The classical propositional and predicate fragment of G-Calculus, namely Gc, is obtained by restricting the number of formulae in the succedent of LK to a single formula C while preserving nearly all other logical and structural rules of LK alongside the addition of two supplementary logical rules: L⊥ and Gem.[31] The reason for the inclusion of the principle L⊥ in Gc is twofold. On one hand, it ensures the preservation of its singlesuccedent nature by eliminating the necessity of the rule RW, whereas on the other hand it enhances the system's derivability power. Of course we are fully aware that the rule L⊥ can be derived from the simultaneous presence of the four principles; A ⇒ A, RW, L¬ and L∧:[32]

---

[30] There are many more sequent-based systems in Indrzejczak's survey but we do not have sufficient space to discuss all of them in this paper. Therefore we will restrict ourselves to the systems that have been mentioned. However, we should state that Indrzejczak concludes his study by noting that nonstandard sequent calculi provide valuable insights into proof theory and logic formalisation. While Gentzen's framework remains the dominant paradigm, alternative sequent calculi offer unique advantages for specific applications such as automated reasoning, modal logic and substructural logics. The study highlights that the choice of a sequent calculus depends on the balance between inferential power, proof-search efficiency and structural simplicity. While rule-based systems (Gentzen's type) are well-suited for theoretical proof analysis, structural-based systems (Hertz's type) offer a more declarative representation of logical principles. Mixed-type systems attempt to integrate the best of both worlds by adapting sequent frameworks to various logics and reasoning methodologies. Thus Indrzejczak's survey highlights the importance of exploring nonstandard sequent calculi as viable alternatives to traditional proof systems paving the way for new developments in logical formalisation and proof theory.

[31] For acknowledging the distinction between G-Calculus and the system Gc, see footnote 4.

$$
\frac{
\frac{
\frac{
\frac{
\dfrac{A \Rightarrow A}{A \Rightarrow A,C}RW
}{A,\neg A \Rightarrow C}L\neg
}{A \wedge \neg A \Rightarrow C}L\wedge
}{\bot \Rightarrow C}L\bot
}{}
$$

Apart from the rule L⊥, the reason for the presence of the principle Gem and not its atomic counterpart Gem-at in Gc, is to fully preserve it as a complete classical system. While Gem-at suffices for extending intuitionistic systems with atomic-level classical reasoning, it remains fundamentally limited in scope by preventing the sequent calculus from achieving the full derivability of classical tautologies involving arbitrary formulae. Gem-at enforces LEM solely for atomic propositions, which means that for complex formulae composed of logical constants such as ∧, ∨ and →, the classical principle must still be simulated or reconstructed indirectly. This severely impairs the proof-theoretic ambition of G-Calculus to serve as a foundational classical system. In contrast, Gem grants LEM at the level of arbitrary formulae, thus enabling direct derivations of fully classical sequents without the need for additional translation or schema. Accordingly, Gem secures the expressive range necessary for capturing the totality of classical entailments, therefore ensuring that G-Calculus meets both its formal and philosophical objective of being a genuinely absolute classical singlesuccedent sequent calculus.

Furthermore, the adoption of Gem rather than Gem-at brings about substantial proof-theoretic advantages particularly concerning the structural symmetry between left and right rules. Since G-Calculus (Gc) includes classical rules such as L¬ and R¬ which manipulate arbitrary formulae rather than restricting themselves to atomic instances, it is essential for the calculus to possess a principle that matches this generality. Gem, by applying LEM universally, preserves the isomorphic relations between left and right negation rules, and by extension, supports the seamless application of cut across arbitrary formulae. Without Gem, derivations involving non-atomic negations would encounter an asymmetry, undermining the intended classical harmony of the system. Moreover, the availability of Gem fosters uniformity in proof search and guarantees that complex formulae can be resolved classically without resorting to reductions to atomic cases which would otherwise fragment the proof-theoretic architecture of the calculus. As a result, the presence of Gem is crucial for maintaining the structural consistency and classical identity of G-Calculus.

---

[32] Two small but important notes should be mentioned. In proof theory, formulae located on the left of the ⇒ which are formally linked via commas e.g. A,¬A ⇒ in an informal setting, that is to say in a logical environment, are conjunctively read as A ∧ ¬A, plus the logical contradictory state A ∧ ¬A, is replaced with the zero-place formal operator ⊥. This means in a proof instead of writing A,¬A ⇒ C, we just inscribe ⊥ ⇒ C.

Finally, from a broader proof-theoretic perspective, Gem enhances the calculus' ability to preserve critical properties such as invertibility of rules and admissibility of structural transformations across arbitrary formulae. By enabling LEM for all formulae, Gem ensures that derivations can proceed with uniform logical steps, without the need for additional meta-level interventions to bridge the gap between atomic and compound formulae. This not only restructures the construction of proofs but also upholds the subformula discipline more robustly by allowing classical reasoning to operate within the full language of the system rather than fragmenting it into atomic and non-atomic layers. In this way, Gem plays a pivotal role in guaranteeing that G-Calculus is not merely an intuitionistic calculus with classical fragments but a fully classical singlesuccedent sequent calculus in its own right, capable of capturing the entirety of classical logic both in structure and in proof-theoretic behaviour. Subsequently, the following formalisms constitute Gc's axiom, logical and structural rules in their embryonic form for the propositional and predicate calculus:

**Gc**

**Axiom**

$A \Rightarrow A$

**Logical Rules**

$$\frac{A,\Gamma \Rightarrow C}{A \wedge B,\Gamma \Rightarrow C}L\wedge \qquad \frac{B,\Gamma \Rightarrow C}{A \wedge B,\Gamma \Rightarrow C}L\wedge \qquad \frac{\Gamma \Rightarrow A \quad \Gamma \Rightarrow B}{\Gamma \Rightarrow A \wedge B}R\wedge$$

$$\frac{A,\Gamma \Rightarrow C \quad B,\Gamma \Rightarrow C}{A \vee B,\Gamma \Rightarrow C}L\vee \qquad \frac{\Gamma \Rightarrow A}{\Gamma \Rightarrow A \vee B}R\vee \qquad \frac{\Gamma \Rightarrow B}{\Gamma \Rightarrow A \vee B}R\vee$$

$$\frac{\Gamma \Rightarrow A \quad B.\Gamma \Rightarrow C}{A \rightarrow B.\Gamma \Rightarrow C}L\rightarrow \qquad \frac{A.\Gamma \Rightarrow B}{\Gamma \Rightarrow A \rightarrow B}R\rightarrow$$

$$\frac{}{\bot \Rightarrow C}L\bot$$

$$\Gamma \Rightarrow A \qquad\qquad A,\Gamma \Rightarrow \bot \qquad\qquad A.\Gamma \Rightarrow C \qquad\qquad \neg A.\Gamma \Rightarrow C$$

$$\frac{\qquad\qquad}{\neg A,\Gamma \Rightarrow \bot}L\neg \qquad \frac{\qquad\qquad}{\Gamma \Rightarrow \neg A}R\neg \qquad \frac{\qquad\qquad\qquad\qquad}{\Gamma \Rightarrow C}Gem$$

$$\frac{A(x/t),\Gamma \Rightarrow C}{\forall xAx,\Gamma \Rightarrow C}L\forall \qquad \frac{\Gamma \Rightarrow A(x/y)}{\Gamma \Rightarrow \forall xAx}R\forall$$

$$\frac{A(x/y),\Gamma \Rightarrow C}{\exists xAx,\Gamma \Rightarrow C}L\exists \qquad \frac{\Gamma \Rightarrow A(x/t)}{\Gamma \Rightarrow \exists xAx}R\exists$$

**Structural Rules**

$$\frac{\Gamma \Rightarrow C}{A,\Gamma \Rightarrow C}LW \qquad \frac{A.A.\Gamma \Rightarrow C}{A.\Gamma \Rightarrow C}LC \qquad \frac{\Gamma \Rightarrow A \quad A,\Gamma \Rightarrow C}{\Gamma \Rightarrow C}Cut$$

However, in what follows we use our knowledge of recent discoveries in structural proof theory to modernise the above calculus by eliminating a number of proof-theoretic issues and propose a number of formal enhancements.

**Axiom**

Instead of employing the reflexive sequent A $\Rightarrow$ A as our axiom or initial sequent, we henceforth employ the sequent A,$\Gamma$ $\Rightarrow$ A. If we wish, this approach allows us to eliminate the structural rule of weakening from our system. But since G-Calculus has been primarily formulated for classical purposes, therefore, we shall retain weakening as part of our structural arsenals. The inclusion of weakening within our framework ensures that our proof system sufficiently remains expressive to capture the full range of classical proof-theoretic activities. Classical logic is characterised by the principle that the validity of a proof does not necessarily depend on the minimality of its assumptions. This means a conclusion remains valid regardless of whether supplementary i.e. potentially superfluous assumptions are present. This principle is fundamental to several metatheoretical results in proof theory including cut-elimination and interpolation theorems.[33] Equally , from a model-theoretic outlook, weakening corresponds to the principle that truth is monotonic in classical semantics i.e. if a statement is true in a given model it remains true regardless of whether additional premises are introduced. This supports the classical conception of logical consequence

---

[33] From a natural deduction perspective, the retention of weakening allows for a more seamless transition from a sequent calculus machinery to a corresponding natural deduction system. In classical natural deduction, assumptions can be freely introduced and later discharged via rules such as implication introduction or negation introduction. The structural rule of weakening in sequent calculus mirrors this flexibility by allowing the introduction of assumptions without restriction. If we were to remove weakening, we would risk introducing an implicit restriction on natural deduction that could interfere with the formulation of elimination rules and classical principles such as LEM.

wherein an argument remains sound regardless of any expansion of the premise set. Thus, the inclusion of weakening ensures that our proof-theoretic and model-theoretic perspectives remain harmonised.[34]

## Exhibiting Context

In sequent calculus systems, in terms of displaying contexts such as $\Gamma$ or $\Delta$, we should state that from a proof analysis point of view shared and independent context styles are equivalent because each style can be derived from the other one. The following two examples demonstrate the rule R∧ once as shared and then as independent context:

$$\frac{\Gamma \Rightarrow A \quad \Gamma \Rightarrow B}{\Gamma \Rightarrow A \wedge B} R\wedge \qquad\qquad \frac{\Gamma \Rightarrow A \quad \Delta \Rightarrow B}{\Gamma, \Delta \Rightarrow A \wedge B} R\wedge$$

However, in formulating G-Calculus we will use a context sharing formalism because when it comes to performing proof search, context sharing and context independent are not alike. This is because, if sequent calculus rules are employed to find a first-root derivation from the endsequent, subsequently the use of context independent rules brings about an explosion of the combinatorial possibilities of dividing the context in the conclusion between the premises of two-premise rules. Conversely, with regards to shared context, the premises are distinctively decided once the principal formula of the desired sequent is selected.

## Invertibility

Apart from the fact that in every sequent calculus system possessing an invertible structure is always desirable, given that in an invertible calculus a logical rule is invertible if the derivability of a sequent in the form of its succedent ensures the derivability of the corresponding premise or premises, for this reason it is demonstrated in the following paragraphs that an invertible calculus such as our proposed G-Calculus permits the simultaneous introduction of left and right inversion rules. This enables us to put forward fresh and authoritative supplementary logical rules. These rules are designated as left-inversion (LI) and right-inversion (RI) rules wherein the use of the constant in question is explicitly denoted. However, given that G-Calculus is based on the system LK, consequently before listing G-Calculus' LI and RI rules and announcing them as the new logical principles of the system, we must first identify those LK rules that are not invertible. As a result, consider the following two LK's L∧ rules:

$$A, \Gamma \Rightarrow \Delta \qquad\qquad B, \Gamma \Rightarrow \Delta$$

---

[34] Alternative logical frameworks that exclude weakening such as relevance logic and certain systems of linear logic demonstrate that inferential expressivity need not be compromised by its removal. Such systems provide a finer-grained analysis of inferential dependencies by ensuring that every assumption made within a derivation plays a necessary role in producing the conclusion. While these considerations are valuable in certain logical contexts, they do not align with the primary objectives of G-Calculus which aims to serve as a classical system.

$$\frac{}{A \wedge B,\Gamma \Rightarrow \Delta} L\wedge \qquad \frac{}{A \wedge B,\Gamma \Rightarrow \Delta} L\wedge$$

or their developing singlesuccedent adaptation in Gc:

$$\frac{A,\Gamma \Rightarrow C}{A \wedge B,\Gamma \Rightarrow C} L\wedge \qquad \frac{B,\Gamma \Rightarrow C}{A \wedge B,\Gamma \Rightarrow C} L\wedge$$

It is evident that none of the above forms are invertible since the implicative relations A ⊢ A ∧ B and B ⊢ A ∧ B are not logically valid. Nevertheless, in order to convert the above proofs into invertible forms, we are required to apply the structural rule of LW to the two L∧ rules. To prove this point, let us weaken the sequents A,Γ ⇒ C and B,Γ ⇒ C in order to obtain the following formalism:

$$\frac{\dfrac{A,\Gamma \Rightarrow C}{A,B,\Gamma \Rightarrow C} LW}{A \wedge B,\Gamma \Rightarrow C} L\wedge \qquad \frac{\dfrac{B,\Gamma \Rightarrow C}{A,B,\Gamma \Rightarrow C} LW}{A \wedge B,\Gamma \Rightarrow C} L\wedge$$

But we can leave out this additional step of weakening by implementing a single invertible L∧ rule and eradicate the two uninvertible L∧ rules. Therefore, from now on in presenting Gc's L∧ rule, instead of using the two uninvertible single-active formula L∧ rules we implement the double-active formula invertible rule:

$$\frac{A,B,\Gamma \Rightarrow C}{A \wedge B,\Gamma \Rightarrow C} L\wedge$$

Subsequently, we can reveal Gc's LI and RI formulations in the following manner:[35]

**LI∧ & RI∧**

$$\frac{\dfrac{A,B,\Gamma \Rightarrow C}{A \wedge B,\Gamma \Rightarrow C} L\wedge}{A,B,\Gamma \Rightarrow C} LI\wedge \qquad \frac{\dfrac{\Gamma \Rightarrow A \quad \Gamma \Rightarrow B}{\Gamma \Rightarrow A \wedge B} R\wedge}{\Gamma \Rightarrow A} RI\wedge \qquad \frac{\dfrac{\Gamma \Rightarrow A \quad \Gamma \Rightarrow B}{\Gamma \Rightarrow A \wedge B} R\wedge}{\Gamma \Rightarrow B} RI\wedge$$

---

[35] Keep in mind that these LI and RI formulations are intended to demonstrate how a left or right inversion rule is formally manufactured. However, when we wish to present the LI and RI rules of each constant, we will display their final forms; namely, we will remove the first line of these formulations and simply exhibit the inversion rule as the resulting inversion-based logical rule.

**LI∨ & RI∨**

$$\frac{A,\Gamma \Rightarrow C \quad B,\Gamma \Rightarrow C}{A \vee B,\Gamma \Rightarrow C} L\vee \qquad \frac{A,\Gamma \Rightarrow C \quad B,\Gamma \Rightarrow C}{A \vee B,\Gamma \Rightarrow C} L\vee$$

$$\frac{A \vee B,\Gamma \Rightarrow C}{A,\Gamma \Rightarrow C} LI\vee \qquad \frac{A \vee B,\Gamma \Rightarrow C}{B,\Gamma \Rightarrow C} LI\vee$$

$$\frac{\Gamma \Rightarrow A}{\Gamma \Rightarrow A \vee B} R\vee \qquad \frac{\Gamma \Rightarrow B}{\Gamma \Rightarrow A \vee B} R\vee$$

$$\frac{\Gamma \Rightarrow A \vee B}{\Gamma \Rightarrow A} RI\vee \qquad \frac{\Gamma \Rightarrow A \vee B}{\Gamma \Rightarrow B} RI\vee$$

**LI→ & RI→**

$$\frac{\Gamma \Rightarrow A \quad B.\Gamma \Rightarrow C}{A \rightarrow B.\Gamma \Rightarrow C} L\rightarrow \qquad \frac{A.\Gamma \Rightarrow B}{\Gamma \Rightarrow A \rightarrow B} R\rightarrow$$

$$\frac{A \rightarrow B.\Gamma \Rightarrow C}{B.\Gamma \Rightarrow C} LI\rightarrow \qquad \frac{\Gamma \Rightarrow A \rightarrow B}{\Gamma \Rightarrow B} RI\rightarrow$$

**LI∀ & RI∀**

$$\frac{A(x/t),\Gamma \Rightarrow C}{\forall xAx,\Gamma \Rightarrow C} L\forall \qquad \frac{\Gamma \Rightarrow A(x/y)}{\Gamma \Rightarrow \forall xAx} R\forall$$

$$\frac{\forall xAx,\Gamma \Rightarrow C}{A(x/t),\Gamma \Rightarrow C} LI\forall \qquad \frac{\Gamma \Rightarrow \forall xAx}{\Gamma \Rightarrow A(x/t)} RI\forall$$

**LI∃ & RI∃**

$$\frac{A(x/y),\Gamma \Rightarrow C}{\exists xAx,\Gamma \Rightarrow C} L\exists \qquad \frac{\Gamma \Rightarrow A(x/t)}{\Gamma \Rightarrow \exists xAx} R\exists$$

$$\frac{\exists xAx,\Gamma \Rightarrow C}{A(x/y),\Gamma \Rightarrow C} LI\exists \qquad \frac{\Gamma \Rightarrow \exists xAx}{\Gamma \Rightarrow A(x/y)} RI\exists$$

We are already aware that the rules L¬ and R¬ are not invertible. This fact can be demonstrated from the following formalism:

$$\Gamma \Rightarrow A \qquad\qquad A,\Gamma \Rightarrow \bot$$

| | **L¬** | | **R¬** |
|---|---|---|---|
| **¬A,Γ ⇒⊥** | | **Γ ⇒¬A** | |
| | **LI¬** | | **RI¬** |
| **Γ ⇒A** | | **A,Γ ⇒⊥** | |

We can easily understand from the above proofs that L¬ and R¬ are not invertible rules since the implicative relation A ⊢ ¬A or the implicative form ¬A ⊢ A is not logically valid. As a result, the two rules LI¬ and RI¬ are not valid rules. Accordingly, in light of the foregoing analysis, it becomes evident that the explicit integration and independent articulation of left and right-inversion rules within G-Calculus not only refines the syntactic architecture of the system but also yields significant proof-theoretic advantages. Chief among these is the preservation of the subformula property particularly through right-inversion rules associated with ∧ (RI∧), ∨ (RI∨) and → (RI→).[36] These rules ensure that every formula introduced in a derivation is a subformula of the endsequent, thus supporting core metatheoretical results such as cut-elimination and the admissibility of derivations devoid of extraneous formulae.[37] Right-inversion rules promote modular proof construction through permitting logical constants in the succedent to be handled independently of the antecedent's complexity.[38] Furthermore, they are especially valuable in the context of automated proof search where goal-directed backward reasoning is preferred. In such contexts, the deterministic nature of right-inversion steps enhances proof tractability and algorithmic efficiency.

Nevertheless, it must be acknowledged that left-inversion rules also contribute meaningful structural benefits particularly in systems that aim to restrict or eliminate contraction. Left-inversion rules make possible the decomposition of complex assumptions in the antecedent by enabling a granular and economically structured derivation without syntactic redundancy. Conversely, ¬ fails to support a valid right-inversion rule, a limitation that significantly undermines the subformula property in proofs involving negated formulae. Since classical sequent systems require auxiliary operators such as ⊥ to manipulate ¬, its treatment becomes inherently non-analytic and indirect. This structural asymmetry highlights the broader significance of right-inversion rules, not only as tools for proof construction but as critical mechanisms that preserve analytic integrity.[39] The distinction between LI and RI rules, therefore, is not merely a matter of formal elegance but a fundamental requirement for sustaining both syntactic discipline and semantic clarity within a classical singlesuccedent sequent calculus such as G-Calculus.[40]

---

[36] However, it must be noted that while the RI rules for the propositional constants ∧, ∨, and → readily satisfy the subformula property, the corresponding RI rule for ∀ warrants careful scrutiny. This is because, in the rule R∀ where one infers ∀xAx from A(x/y) with y as a bound variable not occurring free in A, the subformula property is typically preserved: the formula ∀xAx is a syntactic extension of A(x/y), introducing no new logical connectives or nonlogical terms beyond those present in the original formula. However, in the inversion rule, where one infers A(x/t) from ∀xAx, the introduction of an arbitrary free term t risks violating the subformula property. This is due to the fact that t may introduce new function symbols or terms not originally present in the quantified formula, thus extending the derivation beyond the strict confines of the subformula discipline. Since the subformula property demands that all formulae within an inference are subformulae of the endsequent, the presence of novel terms in the inversion of R∀ disrupts this requirement. A similar difficulty arises in the case of the rule RI∃, where the formula A(x/y) introduced during the inversion can likewise introduce complexities that compromise strict adherence to the subformula property. Moreover, the double-negation elimination rule RI¬¬ also inherently fails to uphold the subformula property, as deriving A from ¬¬A involves introducing a formula not strictly present as a subformula of the initial formula, thus undermining the integrity of the subformula framework. These considerations highlight that not all right-inversion rules uniformly satisfy the subformula property: while RI rules for the propositional constants ∧, ∨, and → typically preserve it, the rules for quantifiers and negation introduce challenges that must be carefully managed within the system.

[37] As we shall see in future works, right-inversion rules are also beneficial for proof normalisation in matching natural deduction systems.

[38] Another major use of the RI rules can be seen when we aim to convert the corresponding natural deduction of G-Calculus into sequent style. In such a conversion, the RI rules will constitute the special elimination rules of the system.

## Conjunction & Disjunction Duality

In classical logic, the two constants ∧ and ∨ display a primary duality relation which is grounded in De Morgan's laws and the structure of Boolean Algebras. This duality arises from the way these constants cooperate with ¬ permitting one to be articulated in terms of the other. De Morgan's laws such as ¬(A ∧ B) ↔ ¬A ∨ ¬B or ¬(A ∨ B) ↔ ¬A ∧ ¬B exhibit that ∧ and ∨ perform as symmetric or mirror images under negation. As a result, the two R∨ rules:

$$\frac{\Gamma \Rightarrow A}{\Gamma \Rightarrow A \vee B} R\vee \qquad\qquad \frac{\Gamma \Rightarrow B}{\Gamma \Rightarrow A \vee B} R\vee$$

are the symmetric images of our two LK-based but single-active formula L∧ rules:

$$\frac{A,\Gamma \Rightarrow C}{A \wedge B,\Gamma \Rightarrow C} L\wedge \qquad\qquad \frac{B,\Gamma \Rightarrow C}{A \wedge B,\Gamma \Rightarrow C} L\wedge$$

However, since we have altered the above single-active formula L∧ rules to a double-active formulae L∧ rule:

$$\frac{A,B,\Gamma \Rightarrow C}{\qquad\qquad} L\wedge$$

---

[39] As we shall see in the subsequent paper which will follow this work as one of its direct proof-theoretic results, a further significance of the RI rules is the fact that these rules in the cases of RI∧, RI∨, RI→, RI¬¬, RI∀, and RI∃, constitute the formal, sequent calculus-based sources of the special elimination rules for the respective constants in natural deduction.

[40] Naturally, the extraction and formulation of the LI and RI rules are not unique proof-theoretic features of G-Calculus alone; they can be formulated within any standard and symmetric sequent calculus, whether singlesuccedent or multisuccedent in structure. Nevertheless, we maintain that these inversion rules, irrespective of the calculus in which they are situated, ought to be identified and subjected to systematic proof-theoretic analysis.

$A \wedge B, \Gamma \Rightarrow C$

therefore, apart from its classical advantages which will be explained below, in order to preserve the sequent calculus duality property between $\wedge$ and $\vee$, accordingly, we are equally required to change our two R$\vee$ rules from a single-active formula rule to a single but double-based active-formulae rule like:

$$\Gamma \Rightarrow A, B$$
$$\underline{\qquad\qquad} R\vee$$
$$\Gamma \Rightarrow A \vee B$$

But prior to elucidating the proof-theoretic significance of the above R$\vee$ rule within a classical sequent calculus apparatus such as G-Calculus (or Gc), we must first defend ourselves against the possible accusation stating that: 'This rule is not a singlesuccedent rule but rather a multisuccedent principle for the reason that the premise $\Gamma \Rightarrow A, B$, possesses more than one formula in the succedent.' However, in response to this allegation through a series of arguments we declare that a closer proof-theoretic examination reveals that this rule is not multisuccedent in its logical character. Rather, it belongs to a singlesuccedent system. This becomes apparent once we scrutinise the structure, semantic goal and proof-theoretic constraints of this principle particularly within the broader aims of proof theory. First, in classical and intuitionistic sequent calculi, sequents in the form $\Gamma \Rightarrow \Delta$ are regularly interpreted as: if all the formulae in $\Gamma$ are true, then at least one formula in $\Delta$ must be true. In multisuccedent systems like Gentzen's LK, the succedent may contain multiple formulae, permitting rules that control or retain multiple possible conclusions. But in singlesuccedent schemes such as Gentzen's LJ or adapted classical systems, for technical reasons, the succedent is limited to a single formula. This design is not merely syntactic. It reflects a significant conceptual constraint, namely the conclusion must consist in a specific, determinate outcome and not merely a disjunction of possibilities. The conclusion of the R$\vee$ rule $\Gamma \Rightarrow A, B \vdash \Gamma \Rightarrow A \vee B$ consists of a single, compound formula. The derivation leads to the assertion that A or B follows from $\Gamma$. In a singlesuccedent system such a formula represents an permissible output. Subsequently, the question becomes: 'Does the premise $\Gamma \Rightarrow A, B$ compromise the singlesuccedent nature of the rule?' We argue the answer is no. This is because, despite the appearance of multisuccedent formulae in the premise, this is not an expression of a general multisuccedent framework. Instead, it represents an intermediate syntactic step designed to mirror the semantic notion that $A \vee B$ is true if either A or B (or both) are true. The multiplicity in the premise enables the syntactic expression of this disjunctive truth condition without committing the system to multisuccedent reasoning in the conclusion. Furthermore, the multisuccedent form of this R$\vee$ rule is $\Gamma \Rightarrow \Delta, A, B \vdash \Gamma \Rightarrow \Delta, A \vee B$ where the context $\Delta$ has made this derivation a multisuccedent derivation. In this multisuccedent proof, unlike the previous singlesuccedent one where the truth of $A \vee B$ was because of the truth of A or B (or both), in the multisuccedent shape the succedent could be true while A and B are both false. This is because, if at least one formula in $\Delta$ is true, subsequently the whole succedent $\Delta, A \vee B$ will be true.

Second, proof-theoretically speaking, the distinction between singlesuccedent and multisuccedent systems is not simply a matter of which sequents are permitted, but of how logical and structural rules cooperate with the form of sequents. A system is proof-theoretically singlesuccedent if all conclusions of logical rules are singlesuccedent sequents and if the system maintains closure under derivations in which only a single formula ever appears in the succedent of the final result. For example, in the case of the R $\vee$ rule $\Gamma \Rightarrow A,B$ ⊢ $\Gamma \Rightarrow A \vee B$, although the premise contains two formulae in the succedent, it functions as an internal step within a derivation whose goal is to introduce a single formula i.e. $A \vee B$. In other words, the derivation is only valid if this single formula is the final outcome. At no point does the rule permit the preservation of both A and B in the succedent of the conclusion. The presence of both A and B in the premise serves to demonstrate that from $\Gamma$, it is possible to derive either one or the other (or both) justifying the assertion of their disjunction. Conversely, a genuinely multisuccedent principle would permit a conclusion like $\Gamma \Rightarrow A,B$ plus further rules would be available to manipulate multisuccedent freely e.g. to eliminate A while preserving B or vice versa, or to distribute structural rules across the succedent. However, the R $\vee$ rule does not do this. It collapses the multiplicity of the premise into a single disjunctive formula. Thus, the rule exhibits a convergent behaviour that is moving from potentially multiple candidate outcomes toward a single, compound conclusion. Additionally, a central proof-theoretic criterion for identifying singlesuccedent systems is the admissibility of cut and the possibility of cut-elimination while maintaining singlesuccedent constraints throughout. In a system employing this form of R $\vee$, all derivations culminating in $\Gamma \Rightarrow A \vee B$ can be understood as terminating in a singlesuccedent endsequent like our desired model sequent $\Gamma \Rightarrow C$:

$$\cfrac{\cfrac{\Gamma \Rightarrow A,B}{\Gamma \Rightarrow A \vee B} R \vee \qquad \cfrac{A,\Gamma \Rightarrow C \quad B,\Gamma \Rightarrow C}{A \vee B,\Gamma \Rightarrow C} L \vee}{\Gamma \Rightarrow C} Cut$$

Even if the premises introduce temporary multiformulae in the succedent, the final conclusion adheres to the singlesuccedent format. Cut-elimination procedures can preserve this restriction by demonstrating that the rule is compatible with the proof-theoretic architecture of a singlesuccedent system. Ultimately, let us consider how structural rules behave in relation to the R $\vee$ rule in question. In multisuccedent systems weakening and contraction in the succedent are essential and permissible by permitting one to extend or reduce the set of conclusions. However, this R $\vee$ rule does not rely on succedent weakening or succedent contraction. There is no step in its use that permits substituting $A \vee B$ for A or B or both independently. The rule strictly introduces $A \vee B$ as a compound formula and it does so from a structured premise that syntactically signals the possibility of either disjunct being derivable. In this sense the rule is a compression principle that is, it compacts potential multiconclusion into a single logical compound. Such behaviour characterises singlesuccedent systems which privilege unique conclusions over multiple potential ones. In summary, it is comprehensible that the R $\vee$ rule:

$$\frac{\Gamma \Rightarrow A, B}{\Gamma \Rightarrow A \vee B} R\vee$$

is best understood as a singlesuccedent rule. Regardless of the syntactic appearance of more than one succedent in the premise, the rule's logical and proof-theoretic structure does not support general multisuccedent reasoning. Rather, the rule enables the derivation of a single, compound conclusion i.e. A ∨ B which supports the principles of singlesuccedent systems. From a proof-theoretic perspective, the rule contributes to a sequent calculus framework in which derivations terminate in unique conclusions and structural rules are appropriately constrained. As a result, the rule exemplifies the subtle but crucial distinction between syntactic appearance and proof-theoretic commitment in the theory of sequents. However, getting back to the importance of this R∨ rule for the calculus Gc (G-Calculus), we should state that its inversion instance is:[41]

$$\frac{\dfrac{\Gamma \Rightarrow A, B}{\Gamma \Rightarrow A \vee B} R\vee}{\Gamma \Rightarrow A, B} RI\vee$$

Arguably, the above rule should be considered as the central classical form of the sequent calculus' R∨ rule. This is because, the key distinction between a classical system and a non-classical one is that the classical classification permits us to derive LEM:

---

[41] A further point that needs to be mentioned is the status of the inversion instance of the R∨ rule, namely $\Gamma \Rightarrow A \vee B \vdash \Gamma \Rightarrow A, B$. At first glance, one might suspect that this instance introduces a multisuccedent framework, since it reiterates a sequent in which two formulae appear in the succedent. However, upon closer examination, it becomes clear that this inversion instance preserves the singlesuccedent character of the original rule rather than violating it. The crucial observation lies in the fact that this RI∨ instance is not an independent rule acting upon a generalised multisuccedent base, but is in fact the inversion of the original premise sequent $\Gamma \Rightarrow A, B$. As such, its purpose is not to introduce new structural forms but rather to recover, for meta-theoretical purposes, the antecedent stage of the R∨ rule in a way that demonstrates the semantic equivalence and inferential symmetry of the derivation. As was stated, the proof-theoretic role of inversion rules is to elucidate how one may reconstruct the premises of a derivation from its conclusion. In this case, since the original R∨ rule transitions from $\Gamma \Rightarrow A, B$ to $\Gamma \Rightarrow A \vee B$, the inversion instance RI∨ effectively verifies that any derivation of $\Gamma \Rightarrow A \vee B$ must, in principle, rely on a background justification akin to $\Gamma \Rightarrow A, B$. This operation, however, does not introduce genuinely new multiplicity into the calculus. The structure remains faithful to the proof-theoretic behaviour of singlesuccedent systems, since the ultimate aim continues to be the derivation and validation of the single, compound conclusion A ∨ B. Moreover, it should be emphasised that no multisuccedent derivation has occurred in this process. The sequent $\Gamma \Rightarrow A, B$ is not treated as a final conclusion but as an internal syntactic expression required for the introduction of disjunction. The inversion rule, far from enabling arbitrary multisuccedent manipulation, simply recapitulates this internal step. The absence of structural rules allowing the elimination, duplication, or permutation of succedent components further reinforces this point. In conclusion, both the original R∨ rule and its inversion RI∨ should be viewed as singlesuccedent rules whose brief syntactic multiplicity serves a confined, semantically transparent role within an otherwise strictly singlesuccedent system.

$$\frac{A \Rightarrow A}{\Rightarrow A, \neg A} R\neg$$

$$\frac{\Rightarrow A, \neg A}{\Rightarrow A \lor \neg A} R\lor$$

where the non-classical system does not allow this derivation. But apart from stating the obvious, we can see from the last line of the above derivation that LEM has been derived via the application of a double-active formulae $R\lor$ rule namely from the simultaneous disjunction of the sequent $\Rightarrow A, \neg A$. However, if we wish to derive LEM via the single-active formula $R\lor$ rule then our derivation could look like:

$$\frac{A \Rightarrow A}{A \Rightarrow A \lor \neg A} R\lor$$

But the problem with the above derivation is that if we believe we have derived LEM through a single-active formula $R\lor$ rule, then LEM can be no longer considered as a proof-theoretic Hilbert-style axiom or in a model-theoretic terms, a logically true theorem. This is because, when we derive LEM in a sequent calculus format in the form of $\Rightarrow A \lor \neg A$, immediately from the logical form of this sequent we will know that the formula which is placed on the right of the $\Rightarrow$ is a Hilbert-style axiom and regardless of the formulae located on the left of the $\Rightarrow$, it is logically true. However, if we recognise the sequent $A \Rightarrow A \lor \neg A$ as the derivation of the law of the excluded middle, but for the reason that LEM has been placed on the right of the $\Rightarrow$ then one can argue this sequent cannot be identified as a proof-theoretic (Hilbert-style) axiom since the left of the $\Rightarrow$ is not empty. Hence, we conclude that from a classical point of view, the core sequent calculus rule $R\lor$ should be employed as a double-active formulae principle and the two single-active formula $R\lor$ rules should be regarded as supplementary $R\lor$ rules of our system. Next, while we altered our $L\land$ and $R\lor$ rules from a single-active formula format to a double-active formulae types, our $R\land$ and $L\lor$ rules still remain as before.

**Left & Right Double Negation Rules**

The final set of rules to be added to the calculus Gc are the two classical rules $L\neg\neg$ and $R\neg\neg$:

$$\frac{A, \Gamma \Rightarrow C}{\neg\neg A, \Gamma \Rightarrow C} L\neg\neg \qquad \frac{\Gamma \Rightarrow A}{\Gamma \Rightarrow \neg\neg A} R\neg\neg$$

and their corresponding two inversion instances:

$$A, \Gamma \Rightarrow C \qquad\qquad \Gamma \Rightarrow A$$

| $\mathbf{L}\neg\neg$ | $\mathbf{R}\neg\neg$ |
|---|---|
| $\neg\neg\mathbf{A}, \Gamma \Rightarrow \mathbf{C}$ | $\Gamma \Rightarrow \neg\neg\mathbf{A}$ |
| $\mathbf{LI}\neg\neg$ | $\mathbf{RI}\neg\neg$ |
| $\mathbf{A}, \Gamma \Rightarrow \mathbf{C}$ | $\Gamma \Rightarrow \mathbf{A}$ |

Although Gentzen (1935) in his LK did not employ these rules and only used L¬ and R¬ to complete his classical sequent calculus, we shall include them in order to solidify G-Calculus as a fully classical system capable of expressing all forms of classical derivations.[42] This means in a classical singlesuccedent sequent calculus system L¬¬ and R¬¬ put forward significant proof-theoretic advantages. They permit for the explicit manipulation of double negation in both the antecedent and succedent thus enriching the expressive and deductive capacities of the system without disrupting its consistency or completeness. In classical logic, the equivalence between a formula and its double negation namely A ↔ ¬¬A is admissible. However, this equivalence must be explicitly managed in a sequent calculus framework. By introducing the two rules L¬¬ and R¬¬ one achieves a syntactic mechanism for introducing or eliminating double negation as needed thus avoiding unnecessary detours in proofs. This capability restructure derivations that otherwise require additional logical manoeuvres to accommodate negated forms. The rule L¬¬ is particularly useful when one needs to apply a premise in the antecedent that appears in a doubly negated form. Without this sequent rule one would need to employ indirect strategies or appeal to derived lemmas to simulate the derivation. But through L¬¬ one can straightforwardly reduce such a doubly negated antecedent to its simpler form. From a structural point of view, this enhances the locality of derivation i.e. each rule acts only on a limited part of the sequent preserving the modularity of the proof. On the other hand, R¬¬ ensures that derivations requiring the double negation of a formula in the succedent do not become artificially complex. In contexts where a formula has been derived but its negated-negation is required (perhaps to match the structure of another formula or to satisfy a later rule application), R¬¬ warrants the system to seamlessly accommodate that need. This is crucial in classical systems where derivability often requires precise syntactic forms and the transformation from A to ¬¬A must be systematised. Together these rules also play a central role in supporting completeness and cut-elimination. They permit double negation to be freely inserted or removed where necessary which is especially important in cut-free proofs. Their admissibility guarantees that all classical tautologies involving double negation can be derived within the system without extending the language or relying on meta-logical arguments. In conclusion, the rules L¬¬ and R¬¬ contribute significantly to the robustness of a classical singlesuccedent sequent calculus such as G-Calculus. They provide essential tools for manipulating negation in a controlled and syntactic manner therefore enhancing proof construction and internal lucidity of the system.[43] Eventually, we re-present the calculus Gc but this time in its modernised form as:

## Gc

---

[42] For Gentzen's original formulation of his LK, see Gentzen (1935) cited in Szabo (1969) Page: 83.

[43] As we shall see in future works, the rule R¬¬, through the assistance of its right-inversion rule RI¬¬, enables us to formulate normalisable double negation introduction (DNI) and double negation elimination (DNE) rules within a natural deduction environment.

**Axiom**

**A,Γ ⇒A**

**Logical Rules**

$$\frac{A,B,\Gamma \Rightarrow C}{A \wedge B,\Gamma \Rightarrow C}L\wedge \qquad \frac{\Gamma \Rightarrow A \quad \Gamma \Rightarrow B}{\Gamma \Rightarrow A \wedge B}R\wedge$$

$$\frac{A \wedge B,\Gamma \Rightarrow C}{A,B,\Gamma \Rightarrow C}LI\wedge \qquad \frac{\Gamma \Rightarrow A \wedge B}{\Gamma \Rightarrow A}RI\wedge \qquad \frac{\Gamma \Rightarrow A \wedge B}{\Gamma \Rightarrow B}RI\wedge$$

$$\frac{A,\Gamma \Rightarrow C \quad B,\Gamma \Rightarrow C}{A \vee B,\Gamma \Rightarrow C}L\vee \qquad \frac{\Gamma \Rightarrow A,B}{\Gamma \Rightarrow A \vee B}R\vee$$

$$\frac{A \vee B,\Gamma \Rightarrow C}{A,\Gamma \Rightarrow C}LI\vee \qquad \frac{A \vee B,\Gamma \Rightarrow C}{B,\Gamma \Rightarrow C}LI\vee \qquad \frac{\Gamma \Rightarrow A \vee B}{\Gamma \Rightarrow A,B}RI\vee$$

$$\frac{\Gamma \Rightarrow A \quad B.\Gamma \Rightarrow C}{A \rightarrow B.\Gamma \Rightarrow C}L\rightarrow \qquad \frac{A.\Gamma \Rightarrow B}{\Gamma \Rightarrow A \rightarrow B}R\rightarrow$$

$$\frac{A \rightarrow B.\Gamma \Rightarrow C}{B.\Gamma \Rightarrow C}LI\rightarrow \qquad \frac{\Gamma \Rightarrow A \rightarrow B}{\Gamma \Rightarrow B}RI\rightarrow$$

$$\frac{}{\bot,\Gamma \Rightarrow C}L\bot$$

$$\frac{A,\Gamma \Rightarrow C}{\neg\neg A,\Gamma \Rightarrow C}L\neg\neg \qquad \frac{\Gamma \Rightarrow A}{\Gamma \Rightarrow \neg\neg A}R\neg\neg$$

$$\frac{\neg\neg A,\Gamma \Rightarrow C}{A,\Gamma \Rightarrow C}LI\neg\neg \qquad \frac{\Gamma \Rightarrow \neg\neg A}{\Gamma \Rightarrow A}RI\neg\neg$$

$$\frac{\Gamma \Rightarrow A}{\quad}L\neg \qquad \frac{A,\Gamma \Rightarrow \bot}{\quad}R\neg \qquad \frac{A.\Gamma \Rightarrow C \qquad \neg A.\Gamma \Rightarrow C}{\quad}Gem$$

$$\dfrac{\neg A,\Gamma \Rightarrow \bot}{\qquad} \qquad\qquad \dfrac{\Gamma \Rightarrow \neg A}{\qquad} \qquad\qquad \dfrac{\Gamma \Rightarrow C}{\qquad}$$

$$\dfrac{A(x/t),\Gamma \Rightarrow C}{\forall xAx,\Gamma \Rightarrow C}\,L\forall \qquad \dfrac{\Gamma \Rightarrow A(x/y)}{\Gamma \Rightarrow \forall xAx}\,R\forall$$

$$\dfrac{\forall xAx,\Gamma \Rightarrow C}{A(x/t),\Gamma \Rightarrow C}\,LI\forall \qquad \dfrac{\Gamma \Rightarrow \forall xAx}{\Gamma \Rightarrow A(x/t)}\,RI\forall$$

$$\dfrac{A(x/y),\Gamma \Rightarrow C}{\exists xAx,\Gamma \Rightarrow C}\,L\exists \qquad \dfrac{\Gamma \Rightarrow A(x/t)}{\Gamma \Rightarrow \exists xAx}\,R\exists$$

$$\dfrac{\exists xAx,\Gamma \Rightarrow C}{A(x/y),\Gamma \Rightarrow C}\,LI\exists \qquad \dfrac{\Gamma \Rightarrow \exists xAx}{\Gamma \Rightarrow A(x/y)}\,RI\exists$$

**Structural Rules**

$$\dfrac{\Gamma \Rightarrow C}{A,\Gamma \Rightarrow C}\,LW \qquad \dfrac{A.A.\Gamma \Rightarrow C}{A.\Gamma \Rightarrow C}\,LC \qquad \dfrac{\Gamma \Rightarrow A \quad A,\Gamma \Rightarrow C}{\Gamma \Rightarrow C}\,Cut$$

To present the intuitionistic segment of the G-Calculus namely Gi, we first eliminate the axiom A,Γ ⇒A and then the rules Γ ⇒A,B ⊢ Γ ⇒A ∨B, L¬¬, R¬¬, L¬, R¬ and Gem, plus their inversion instances. Subsequently, we substitute Gc's axiom A,Γ ⇒A with P,Γ ⇒P as well as substituting R∨ rule Γ ⇒A,B ⊢ Γ ⇒A ∨B and its corresponding inversion illustration with the two R∨ rules Γ ⇒A ⊢ Γ ⇒A ∨B and Γ ⇒B ⊢ Γ ⇒A ∨B and their matching inversion rules.[44] Equally, the two rules L¬ and R¬ is substituted by the two following intuitionistic negation rules:[45]

$$\dfrac{\Gamma \Rightarrow A}{A \rightarrow \bot,\Gamma \Rightarrow \bot}\,L\neg^{\rightarrow} \qquad \dfrac{A,\Gamma \Rightarrow \bot}{\Gamma \Rightarrow A \rightarrow \bot}\,R\neg^{\rightarrow}$$

The reason that we have suggested the above rules as the intuitionistic alternatives of L¬ and R¬ is that the intuitionistic understanding of ¬ for a given formula e.g. ¬A is the conditional formula A → ⊥. The following rules are the logical and structural rules of the calculus Gi:

---

[44] In the axiom P,Γ ⇒ P, P refers to atomic formulae.

[45] To demonstrate the intuitionistic form of the rules L¬ and R¬ in the calculus Gi, it is important to note that these rules represent a combination of ¬ and →. Therefore, to highlight this distinction from the original ¬ rules in Gc, we have also used → in superscript following ¬.

**Gi**

**Axiom**

$P,\Gamma \Rightarrow P$

**Logical Rules**

$$\frac{A,B,\Gamma \Rightarrow C}{A \wedge B,\Gamma \Rightarrow C}L\wedge \qquad \frac{\Gamma \Rightarrow A \quad \Gamma \Rightarrow B}{\Gamma \Rightarrow A \wedge B}R\wedge$$

$$\frac{A \wedge B,\Gamma \Rightarrow C}{A,B,\Gamma \Rightarrow C}LI\wedge \qquad \frac{\Gamma \Rightarrow A \wedge B}{\Gamma \Rightarrow A}RI\wedge \qquad \frac{\Gamma \Rightarrow A \wedge B}{\Gamma \Rightarrow B}RI\wedge$$

$$\frac{A,\Gamma \Rightarrow C \quad B,\Gamma \Rightarrow C}{A \vee B,\Gamma \Rightarrow C}L\vee \qquad \frac{\Gamma \Rightarrow A}{\Gamma \Rightarrow A \vee B}R\vee \qquad \frac{\Gamma \Rightarrow B}{\Gamma \Rightarrow A \vee B}R\vee$$

$$\frac{A \vee B,\Gamma \Rightarrow C}{A,\Gamma \Rightarrow C}LI\vee \qquad \frac{A \vee B,\Gamma \Rightarrow C}{B,\Gamma \Rightarrow C}LI\vee$$

$$\frac{\Gamma \Rightarrow A \vee B}{\Gamma \Rightarrow A}RI\vee \qquad \frac{\Gamma \Rightarrow A \vee B}{\Gamma \Rightarrow B}RI\vee$$

$$\frac{\Gamma \Rightarrow A \quad B.\Gamma \Rightarrow C}{A \rightarrow B.\Gamma \Rightarrow C}L\rightarrow \qquad \frac{A.\Gamma \Rightarrow B}{\Gamma \Rightarrow A \rightarrow B}R\rightarrow$$

$$\frac{A \rightarrow B.\Gamma \Rightarrow C}{B.\Gamma \Rightarrow C}LI\rightarrow \qquad \frac{\Gamma \Rightarrow A \rightarrow B}{\Gamma \Rightarrow B}RI\rightarrow$$

$$\frac{}{\perp,\Gamma \Rightarrow C}L\perp$$

$$\frac{\Gamma \Rightarrow A}{A \rightarrow \perp,\Gamma \Rightarrow \perp}L\neg \qquad \frac{A,\Gamma \Rightarrow \perp}{\Gamma \Rightarrow A \rightarrow \perp}R\neg$$

$$\frac{A(x/t),\Gamma \Rightarrow C}{\forall xAx,\Gamma \Rightarrow C}L\forall \qquad \frac{\Gamma \Rightarrow A(x/y)}{\Gamma \Rightarrow \forall xAx}R\forall$$

$$\frac{\forall xAx,\Gamma \Rightarrow C}{A(x/t),\Gamma \Rightarrow C}LI\forall \qquad \frac{\Gamma \Rightarrow \forall xAx}{\Gamma \Rightarrow A(x/t)}RI\forall$$

$$\frac{A(x/y),\Gamma \Rightarrow C}{\exists xAx,\Gamma \Rightarrow C}L\exists \qquad \frac{\Gamma \Rightarrow A(x/t)}{\Gamma \Rightarrow \exists xAx}R\exists$$

$$\frac{\exists xAx,\Gamma \Rightarrow C}{A(x/y),\Gamma \Rightarrow C}LI\exists \qquad \frac{\Gamma \Rightarrow \exists xAx}{\Gamma \Rightarrow A(x/y)}RI\exists$$

## Structural Rules

$$\frac{\Gamma \Rightarrow C}{A,\Gamma \Rightarrow C}LW \qquad \frac{A.A.\Gamma \Rightarrow C}{A.\Gamma \Rightarrow C}LC \qquad \frac{\Gamma \Rightarrow A \quad A,\Gamma \Rightarrow C}{\Gamma \Rightarrow C}Cut$$

The minimal form of G-Calculus namely Gm, is obtained by eliminating the rules L⊥, L¬⃗ and R¬⃗ from Gi because minimal logic does not allow the presence of an empty succedent.[46] This means in Gm we cannot have a derivation containing the sequent $\Gamma \Rightarrow \perp$. Therefore, there must be always a formula present after the ⇒. The calculus Gm can be presented as:

**Gm**

**Axiom**

**P,$\Gamma \Rightarrow$P**

**Logical Rules**

$$\frac{A,B,\Gamma \Rightarrow C}{A \wedge B,\Gamma \Rightarrow C}L\wedge \qquad \frac{\Gamma \Rightarrow A \quad \Gamma \Rightarrow B}{\Gamma \Rightarrow A \wedge B}R\wedge$$

$$\frac{A \wedge B,\Gamma \Rightarrow C}{A \wedge B,\Gamma \Rightarrow C}LI\wedge \qquad \frac{\Gamma \Rightarrow A \wedge B}{}RI\wedge \qquad \frac{\Gamma \Rightarrow A \wedge B}{}RI\wedge$$

---

[46] See Troelstra & Schwichtenberg (2000) Page: 62.

$$\frac{A,B,\Gamma \Rightarrow C}{A,\Gamma \Rightarrow C \quad B,\Gamma \Rightarrow C} \quad \frac{\Gamma \Rightarrow A}{} \quad \frac{\Gamma \Rightarrow B}{}$$

$$\frac{A,\Gamma \Rightarrow C \quad B,\Gamma \Rightarrow C}{A \vee B,\Gamma \Rightarrow C}L\vee \qquad \frac{\Gamma \Rightarrow A}{\Gamma \Rightarrow A \vee B}R\vee \qquad \frac{\Gamma \Rightarrow B}{\Gamma \Rightarrow A \vee B}R\vee$$

$$\frac{A \vee B,\Gamma \Rightarrow C}{A,\Gamma \Rightarrow C}LI\vee \qquad \frac{A \vee B,\Gamma \Rightarrow C}{B,\Gamma \Rightarrow C}LI\vee$$

$$\frac{\Gamma \Rightarrow A \vee B}{\Gamma \Rightarrow A}RI\vee \qquad \frac{\Gamma \Rightarrow A \vee B}{\Gamma \Rightarrow B}RI\vee$$

$$\frac{\Gamma \Rightarrow A \quad B.\Gamma \Rightarrow C}{A \rightarrow B.\Gamma \Rightarrow C}L\rightarrow \qquad \frac{A.\Gamma \Rightarrow B}{\Gamma \Rightarrow A \rightarrow B}R\rightarrow$$

$$\frac{A \rightarrow B.\Gamma \Rightarrow C}{B.\Gamma \Rightarrow C}LI\rightarrow \qquad \frac{\Gamma \Rightarrow A \rightarrow B}{\Gamma \Rightarrow B}RI\rightarrow$$

$$\frac{A(x/t),\Gamma \Rightarrow C}{\forall xAx,\Gamma \Rightarrow C}L\forall \qquad \frac{\Gamma \Rightarrow A(x/y)}{\Gamma \Rightarrow \forall xAx}R\forall$$

$$\frac{\forall xAx,\Gamma \Rightarrow C}{A(x/t),\Gamma \Rightarrow C}LI\forall \qquad \frac{\Gamma \Rightarrow \forall xAx}{\Gamma \Rightarrow A(x/t)}RI\forall$$

$$\frac{A(x/y),\Gamma \Rightarrow C}{\exists xAx,\Gamma \Rightarrow C}L\exists \qquad \frac{\Gamma \Rightarrow A(x/t)}{\Gamma \Rightarrow \exists xAx}R\exists$$

$$\frac{\exists xAx,\Gamma \Rightarrow C}{A(x/y),\Gamma \Rightarrow C}LI\exists \qquad \frac{\Gamma \Rightarrow \exists xAx}{\Gamma \Rightarrow A(x/y)}RI\exists$$

**Structural Rules**

$$\frac{\Gamma \Rightarrow C}{A,\Gamma \Rightarrow C}LW \qquad \frac{A.A.\Gamma \Rightarrow C}{A.\Gamma \Rightarrow C}LC \qquad \frac{\Gamma \Rightarrow A \quad A,\Gamma \Rightarrow C}{\Gamma \Rightarrow C}Cut$$

To conclude this paper, in what follows we first provide a proof for the calculus Gc for deriving the sequent $\Gamma \Rightarrow C$ from the rule L¬ because this rule like other logical rules of Gc has the proof-theoretic potentiality to derive our desired singlesuccedent sequent. Consequently, consider the following proof:

$$
\frac{\dfrac{A,\Gamma \Rightarrow \bot}{\Gamma \Rightarrow \neg A}R\neg \qquad \dfrac{\Gamma \Rightarrow A}{\neg A,\Gamma \Rightarrow \bot}L\neg}{\dfrac{\Gamma \Rightarrow \bot}{\dfrac{}{}} \text{Cut} \qquad \dfrac{}{\bot,\Gamma \Rightarrow C}L\bot }
$$

A,Γ ⇒⊥                    Γ ⇒A
————————R¬        ————————L¬
  Γ ⇒¬A              ¬A,Γ ⇒⊥
——————————————Cut    ——————L⊥
        Γ ⇒⊥              ⊥,Γ ⇒C
—————————————————————————Cut
            Γ ⇒C

While in order to derive the endsequent $\Gamma \Rightarrow C$ from L¬ we have taken a number of extra steps namely the demonstration of the fact that the rules R¬ and L¬ are symmetrically isomorphic plus the use of the rule L⊥ to be capable to cut the operator ⊥ from the sequent $\Gamma \Rightarrow \bot$ with the aim of deriving the arbitrary formula C from the context $\Gamma$, we have not proof-theoretically done anything wrong and have continued to be in our calculus' territory. Second, we exhibit all right and left logical rules of Gc (excluding their corresponding inversion instances) are isomorphic structures because they have been symmetrically assembled.[47] Regard the following proofs:[48]

Γ ⇒A   Γ ⇒B                A,B,Γ ⇒C
——————————R∧        ——————————L∧
  Γ ⇒A ∧B                A ∧B,Γ ⇒C
——————————————————————————Cut
            Γ ⇒C

Γ ⇒A,B                A,Γ ⇒C   B,Γ ⇒C
—————————R∨        ——————————————L∨
Γ ⇒A ∨B                A ∨B,Γ ⇒C
——————————————————————————Cut
            Γ ⇒C

 A.Γ ⇒B                Γ ⇒A   B.Γ ⇒C
——————————R→        ——————————L→
Γ ⇒A → B                A → B.Γ ⇒C
——————————————————————————Cut
            Γ ⇒C

---

[47] The same isomorphic results can be obtained from the calculi Gi and Gm which are subsystems of Gc.
[48] The rules L⊥ and Gem have been leaved out from this presentation since they by default derive the endsequent $\Gamma \Rightarrow C$.

$$\cfrac{\cfrac{\Gamma \Rightarrow A}{\Gamma \Rightarrow \neg\neg A}\ R\neg\neg \qquad \cfrac{A,\Gamma \Rightarrow C}{\neg\neg A,\Gamma \Rightarrow C}\ L\neg\neg}{\Gamma \Rightarrow C}\ Cut$$

$$\cfrac{\cfrac{\cfrac{A,\Gamma \Rightarrow \bot}{\Gamma \Rightarrow \neg A}\ R\neg \qquad \cfrac{\Gamma \Rightarrow A}{\neg A,\Gamma \Rightarrow \bot}\ L\neg}{\Gamma \Rightarrow \bot}\ Cut \qquad \cfrac{}{\bot,\Gamma \Rightarrow C}\ L\bot}{\Gamma \Rightarrow C}\ Cut$$

$$\cfrac{\cfrac{\Gamma \Rightarrow A(x/y)}{\Gamma \Rightarrow \forall x Ax}\ R\forall \qquad \cfrac{A(x/t),\Gamma \Rightarrow C}{\forall x Ax,\Gamma \Rightarrow C}\ L\forall}{\Gamma \Rightarrow C}\ Cut$$

$$\cfrac{\cfrac{\Gamma \Rightarrow A(x/t)}{\Gamma \Rightarrow \exists x Ax}\ R\exists \qquad \cfrac{A(x/y),\Gamma \Rightarrow C}{\exists x Ax,\Gamma \Rightarrow C}\ L\exists}{\Gamma \Rightarrow C}\ Cut$$

Finally, to exhibit the flexibility of G-Calculus, we demonstrate its classical multisuccedent form namely GM-Calculus (GM) as:[49]

**GM**

**Axiom**

$A,\Gamma \Rightarrow \Delta,A$

**Logical Rules**

$$\cfrac{A,B,\Gamma \Rightarrow \Delta}{}\ L\wedge \qquad \cfrac{\Gamma \Rightarrow \Delta,A \quad \Gamma \Rightarrow \Delta,B}{}\ R\wedge$$

---

[49] Although this matter requires a full independent enquiry which we do not have the space to pursue in this work, it may be briefly noted that GM-Calculus (GM) exhibits a number of proof-theoretic improvements over Gentzen's LK. Some of these advantages were elucidated when we undertook a proof-theoretic modernisation of the G-calculus particularly its classical segment Gc.

A ∧B,Γ ⇒Δ             Γ ⇒Δ,A ∧B

$$\frac{A \wedge B,\Gamma \Rightarrow \Delta}{A,B,\Gamma \Rightarrow \Delta}\ LI\wedge \qquad \frac{\Gamma \Rightarrow \Delta,A \wedge B}{\Gamma \Rightarrow \Delta,A}\ RI\wedge \qquad \frac{\Gamma \Rightarrow \Delta,A \wedge B}{\Gamma \Rightarrow \Delta,B}\ RI\wedge$$

$$\frac{A,\Gamma \Rightarrow \Delta \quad B,\Gamma \Rightarrow \Delta}{A \vee B,\Gamma \Rightarrow \Delta}\ L\vee \qquad \frac{\Gamma \Rightarrow \Delta,A,B}{\Gamma \Rightarrow \Delta,A \vee B}\ R\vee$$

$$\frac{A \vee B,\Gamma \Rightarrow \Delta}{A,\Gamma \Rightarrow \Delta}\ LI\vee \qquad \frac{A \vee B,\Gamma \Rightarrow \Delta}{B,\Gamma \Rightarrow \Delta}\ LI\vee \qquad \frac{\Gamma \Rightarrow \Delta,A \vee B}{\Gamma \Rightarrow \Delta,A,B}\ RI\vee$$

$$\frac{\Gamma \Rightarrow \Delta,A \quad B.\Gamma \Rightarrow \Delta}{A \to B.\Gamma \Rightarrow \Delta}\ L\to \qquad \frac{A.\Gamma \Rightarrow \Delta,B}{\Gamma \Rightarrow \Delta,A \to B}\ R\to$$

$$\frac{A \to B.\Gamma \Rightarrow \Delta}{B.\Gamma \Rightarrow \Delta}\ LI\to \qquad \frac{\Gamma \Rightarrow \Delta,A \to B}{\Gamma \Rightarrow \Delta,B}\ RI\to$$

$$\frac{}{\bot,\Gamma \Rightarrow \Delta}\ L\bot$$

$$\frac{A,\Gamma \Rightarrow \Delta}{\neg\neg A,\Gamma \Rightarrow \Delta}\ L\neg\neg \qquad \frac{\Gamma \Rightarrow \Delta,A}{\Gamma \Rightarrow \Delta,\neg\neg A}\ R\neg\neg$$

$$\frac{\neg\neg A,\Gamma \Rightarrow \Delta}{A,\Gamma \Rightarrow \Delta}\ LI\neg\neg \qquad \frac{\Gamma \Rightarrow \Delta,\neg\neg A}{\Gamma \Rightarrow \Delta,A}\ RI\neg\neg$$

$$\frac{\Gamma \Rightarrow \Delta,A}{\neg A,\Gamma \Rightarrow \Delta}\ L\neg \qquad \frac{A,\Gamma \Rightarrow \Delta}{\Gamma \Rightarrow \Delta,\neg A}\ R\neg \qquad \frac{A.\Gamma \Rightarrow \Delta \quad \neg A.\Gamma \Rightarrow \Delta}{\Gamma \Rightarrow \Delta}\ Gem$$

$$\frac{A(x/t),\Gamma \Rightarrow \Delta}{\forall x Ax,\Gamma \Rightarrow \Delta}\ L\forall \qquad \frac{\Gamma \Rightarrow \Delta,A(x/y)}{\Gamma \Rightarrow \Delta,\forall x Ax}\ R\forall$$

$$\frac{\forall x Ax,\Gamma \Rightarrow \Delta}{A(x/t),\Gamma \Rightarrow \Delta}\ LI\forall \qquad \frac{\Gamma \Rightarrow \Delta,\forall x Ax}{\Gamma \Rightarrow \Delta,A(x/t)}\ RI\forall$$

$$\frac{A(x/y), \Gamma \Rightarrow \Delta}{\exists x Ax, \Gamma \Rightarrow \Delta} \, L\exists \qquad \frac{\Gamma \Rightarrow \Delta, A(x/t)}{\Gamma \Rightarrow \Delta, \exists x Ax} \, R\exists$$

$$\frac{\exists x Ax, \Gamma \Rightarrow \Delta}{A(x/y), \Gamma \Rightarrow \Delta} \, LI\exists \qquad \frac{\Gamma \Rightarrow \Delta, \exists x Ax}{\Gamma \Rightarrow \Delta, A(x/y)} \, RI\exists$$

**Structural Rules**

$$\frac{\Gamma \Rightarrow \Delta}{A, \Gamma \Rightarrow \Delta} \, LW \qquad \frac{\Gamma \Rightarrow \Delta}{\Gamma \Rightarrow \Delta, A} \, RW$$

$$\frac{A, A, \Gamma \Rightarrow \Delta}{A, \Gamma \Rightarrow \Delta} \, LC \qquad \frac{\Gamma \Rightarrow \Delta, A, A}{\Gamma \Rightarrow \Delta, A} \, RC$$

$$\frac{\Gamma \Rightarrow \Delta, A \quad A, \Gamma \Rightarrow \Delta}{\Gamma \Rightarrow \Delta} \, Cut$$

# 4 Concluding Remarks

This paper has introduced G-Calculus as a foundational singlesuccedent sequent calculus tailored for classical logic. Unlike Gentzen's LK, which admits multisuccedent sequents, G-Calculus restricts the succedent to a single formula, hence imposing a syntactic discipline that enhances the system's proof-theoretic transparency. This restriction is not a limitation but a constructive constraint because it clarifies derivations, foregrounds structural symmetry and supports the goal of deriving sequents in the precise form $\Gamma \Rightarrow C$. A number of systems surveyed such as G3ip+Gem-at, G0ip+Gem0-at and Smullyan's multisuccedent calculus demonstrated partial merit but ultimately failed to provide the logical and structural completeness required by a rigorous classical singlesuccedent framework. In response, G-Calculus was formulated to include all essential classical rules including L⊥, L¬¬, R¬¬, L¬, R¬ and Gem, whilst preserving key structural rules such as weakening, contraction and cut.

One of G-Calculus' central innovations lies in its integration of left and right inversion rules. These rules enable refined and modular manipulation of logical constants through revealing isomorphic relationships between left and right patterns across both the antecedent and succedent. The reconfiguration of $\wedge$ and $\vee$ rules particularly the adoption of a double-active R$\vee$ rule with a multiformula premise demonstrates that syntactic flexibility can coexist with singlesuccedent integrity. Detailed analysis confirmed that such formulations do not undermine the system's foundational commitments; rather, they serve as transitional syntactic constructs that preserve a semantically singular outcome. Furthermore, the system's support for symmetric cut derivations reinforces its internal unity, since logical rules appear in structurally balanced pairs by enabling systematic reductions and facilitating proof search. These features collectively affirm the subformula property, maintain syntactic locality and open the path to eventual full cut-elimination.

Finally, G-Calculus' structural clarity extends to its treatment of ¬. The inclusion of L¬¬ and R¬¬ provides crucial proof-theoretic tools for manipulating tautologies and supporting reductio arguments. These additions not only restore analytic control over classical reasoning but also illustrate the expressive range of the system. Moreover, G-Calculus includes its intuitionistic and minimal counterparts (Gi and Gm) as well-defined subsystems, further attesting to its foundational strength and adaptability. Future investigations may include the construction of a corresponding natural deduction system, a full metatheoretic proof of cut-elimination and the development of proof-theoretic semantics aligned with G-Calculus' syntactic design. In summary, G-Calculus is not merely a technical artefact but a reorientation of classical proof theory, namely one that demonstrates how logical completeness, analytic elegance and structural constraint can be harmoniously integrated in a singlesuccedent sequent framework.

# References

Avigad, J. 2004. "Forcing in Proof Theory", *The Bulletin of Symbolic Logic.* Vol. 10. No. 3. pp. 305–333.

Bimbo, K. 2015. *Proof Theory: Sequent Calculi and Related Formalisms.* Florida: CRC Press.

Boričić, B. R. 1986. "A Cut-Free Gentzen-Type System for the Logic of the Weak Law of Excluded Middle", *Studia Logica: An International Journal for Symbolic Logic.* Vol. 45. No. 1. pp. 39–53.

Boričić, B. R. 1988. "A Note on Sequent Calculi Intermediate between LJ and LK", *Studia Logica: An International Journal for Symbolic Logic.* Vol. 47. No. 2. pp. 151–157.

Cellucci, C. 1985. "Proof Theory and Complexity", *Synthesis.* Vol. 62. No. 2. pp. 173–189.

Curry, H. B. 1963. *Foundations of Mathematical Logic.* New York: McGraw-Hill.

Došen, K. 1980. *Logical Constants, an Essay in Proof Theory (PhD Thesis).* University of Oxford.

Došen, K. & Zoran, P. 2007. *Proof-Theoretical Coherence.* London: College Publications.

Dragalin, A. G. 1985. *Mathematical Intuitionism: Introduction to Proof Theory.* Michigan: American Mathematical Society.

Dyckhoff, R. & Negri, S. 2000. "Admissibility of Structural Rules for Contraction-Free Systems of Intuitionistic Logic", *The Journal of Symbolic Logic.* Vol. 65. No. 4. pp. 1499–1518.

Feferman, S. 1975. "Does Reductive Proof Theory Have a Viable Rationale", *Erkenntnis, Concept of Reduction in Logic and Philosophy.* Vol. 53. No. 1/2. pp. 63–96.

Feferman, S. & Seig, W. (eds.) 2010. *Proofs, Categories and Computations: Essays in Honor of Grigori Mints.* College Publications.

Hasenjaeger, G. 1972. *Introduction to the Basic Concepts and Problems of Modern Logic.* Dordrecht: Reidel.

Helman, G. 1983. "An Interpretation of Classical Proofs", *Journal of Philosophical Logic.* Vol. 12. No. 1. pp. 39–71.

Hermes, H. 1963. *Einführung in die Mathematische Logik.* Stuttgart: Teubner.

Hertz, P. 1929. "Über Axiomensysteme für Beliebige Satzsysteme", *Mathematische Annalen.* Vol. 101. pp. 457–514.

Heyting, A. 1971. *Intuitionism: An Introduction.* Amsterdam: North-Holland Publishing Company.

Hilbert, D. & Ackermann, W. 1928. Hammond, L., Leckie, G., Steinhardt, F. (trans.); Luce, R.E. (ed.) *Principles of Mathematical Logic.* New York: Chelsea Publishing Company.

Hsiung, M. 2008. "An Intuitionistic Characterization of Classical Logic", *Journal of Philosophical Logic.* Vol. 37. No. 4. pp. 299–317.

Indrzejczak, A. 2009. "Suszko's Contribution to the Theory of Nonaxiomatic Proof Systems", *Bulletin of the Section of Logic.* Vol. 38. No. 3–4. pp. 151–162.

Indrzejczak, A. 2010. *Natural Deduction, Hybrid Systems and Modal Logics.* Berlin: Springer.

Indrzejczak, A. 2014. "A Survey of Nonstandard Sequent Calculi", *Studia Logica: An International Journal for Symbolic Logic.* Vol. 102. No. 6. pp. 1295–1322.

Jacobsen, F. K. & Villadsen, J. 2024. "A Formalization of Sequent Calculus for Classical Implicational Logic", In: *Proceedings of the Isabelle Workshop 2024.* DTU Technical University.

Kahle, R. & Rathjen, M. 2015. (eds.) *Gentzen's Centenary: The Quest for Consistency.* Springer.

Kleene, S. C. 1952. *Introduction to Metamathematics.* Amsterdam: North-Holland.

Kleene, S. C. 1967. *Mathematical Logic.* New York: Wiley.

Koslow, A. 2007. "Structuralist Logic: Implications, Inferences, and Consequences!", *Logica Universalis.* Vol. 1. pp. 167–181.

Kreisel, G. 1968. "A Survey of Proof Theory", *The Journal of Symbolic Logic.* Vol. 33. No. 3. pp. 321–388.

Krivtsov, V. N. 2010. "An Intuitionistic Completeness Theorem for Classical Predicate Logic", *Studia Logica: An International Journal for Symbolic Logic.* Vol. 96. No. 1. pp. 109–115.

Massaioli, F. 2023. "On the Semantics of Proofs in Classical Sequent Calculus", *arXiv preprint arXiv:2307.16594.*

Mints, G. 1991. "Proof Theory in the USSR 1925–1969", *The Journal of Symbolic Logic.* Vol. 56. No. 2. pp. 385–424.

Mints, G. 2002. *A Short Introduction to Intuitionistic Logic.* New York: Kluwer Academic Publishers.

Negri, S. & Von Plato, J. 1998. "Cut Elimination in the Presence of Axioms", *The Bulletin of Symbolic Logic.* Vol. 4. No. 4. pp. 418–435.

Negri, S. & Von Plato, J. 2001. "Sequent Calculus in Natural Deduction Style", *The Journal of Symbolic Logic.* Vol. 66. No. 4. pp. 1803–1816.

Negri, S. & Von Plato, J. 2001. *Structural Proof Theory.* Cambridge: Cambridge University Press.

Negri, S. & Von Plato, J. 2011. *Proof Analysis: A Contribution to Hilbert's Last Problem.* Cambridge: Cambridge University Press.

Pelletier, F. J. & Hazen, A. P. 2012. "A History of Natural Deduction", In: *Handbook of the History of Logic.* Vol. 11. pp. 341–414.

Pereira, L. C., Haeusler, E. H. & De Paiva, V. (eds.) 2014. *Advances in Natural Deduction: A Celebration of Dag Prawitz.* Dordrecht: Springer Science and Business Media.

Pohlers, W. 1996. "Pure Proof Theory Aims, Methods and Results", *The Bulletin of Symbolic Logic.* Vol. 2. No. 2. pp. 159–188.

Prawitz, D. 1965. *Natural Deduction: A Proof-Theoretical Study.* Stockholm: Almqvist & Wiksell.

Prawitz, D. 1971. "Ideas and Results in Proof Theory", *Journal of Symbolic Logic.* Vol. 40. No. 2. pp. 232–305.

Prawitz, D. 1974. "On the Idea of a General Proof Theory", *Synthesis.* Vol. 27. No. 1/2. pp. 63–77.

Raftey, J. G. 2006. "Correspondence Between Gentzen and Hilbert Systems", *The Journal of Symbolic Logic.* Vol. 71. No. 3. pp. 903–957.

Schindler, R. (ed.) 2010. *Ways of Proof Theory.* Berlin: De Gruyter.

Schwichtenberg, H. & Wainer, S. S. 2011. *Proofs and Computations.* Cambridge: Cambridge University Press.

Schütte, K. 1977. *Proof Theory.* Berlin: Springer.

Sieg, W. 1984. "Foundations for Analysis and Proof Theory", *Synthesis.* Vol. 60. No. 2. pp. 159–200.

Smiley, T. & Shoesmith, D. J. (ed.) 1978. *Multiple-Conclusion Logic.* Cambridge: Cambridge University Press.

Smullyan, R. 1968. *First-Order Logic.* Berlin: Springer.

Smullyan, R. 1995. *First Order Logic.* Mineola, New York: Dover Publications.

Szabo, M. E. (ed.) 1969. *The Collected Papers of Gerhard Gentzen.* Amsterdam: North-Holland Publishing Company.

Tait, W. W. 1968. "Normal Derivability in Classical Logic", In: *The Syntax and Semantics of Infinitary Languages.* LNM 72.

Tait, W. W. 2005. *The Provenance of Pure Reason: Essays in the Philosophy of Mathematics and its History.* Oxford: Oxford University Press.

Tait, W. W. 2006. "Gödel's Correspondence on Proof Theory and Constructive Mathematics", *Philosophia Mathematica.* Vol. 14. No. 1. pp. 76–111.

Tait, W. W. 2006. "Proof-Theoretic Semantics for Classical Mathematics", *Synthesis.* Vol. 148. No. 3. pp. 603–622.

Takeuti, G. 1985. "Proof Theory and Set Theory", *Synthesis.* Vol. 62. No. 2. pp. 255–263.

Toyooka, M. & Sano, K. 2021. "Analytic Multi-Succedent Sequent Calculus for Combining Intuitionistic and Classical Propositional Logic", *Proceedings of ICLA 2021: 9th Indian Conference on Logic and Its Applications.* pp. 128–133.

Toyooka, M. & Sano, K. 2022. "Combining First-Order Classical and Intuitionistic Logic: A Multi-Succedent Sequent Calculus G(FOC+J) with Cut-Elimination", *arXiv preprint arXiv:2204.06723.*

Troelstra, A. S. & Schwichtenberg, H. 2000. *Basic Proof Theory.* Cambridge: Cambridge University Press.

Van Dalen, D. 2008. *Logic and Structure.* Berlin: Springer.

Van Heijenoort, J. (ed.) 1967. *From Frege to Gödel: A Source Book in Mathematical Logic.* Cambridge, Mass: Harvard University Press.

Von Plato, J. 2003. "Rereading Gentzen", *Synthesis.* Vol. 137. No. 1/2. pp. 195–209.

Von Plato, J. 2008. "Gentzen's Proof of Normalization for Natural Deduction", *The Bulletin of Symbolic Logic.* Vol. 14. No. 2. pp. 240–257.

Von Plato, J. 2012. "Gentzen's Proof Systems: Byproducts in a Work of Genius", *The Bulletin of Symbolic Logic.* Vol. 18. No. 3. pp. 313–367.

Von Plato, J. 2014. "Generality and Existence: Quantificational Logic in Historical Perspective", *The Bulletin of Symbolic Logic.* Vol. 20. No. 4. pp. 417–448.